\documentclass{article}

\usepackage{amssymb}
\usepackage{latexsym}\usepackage{booktabs}

\usepackage{url}
\usepackage{xcolor}
\definecolor{newcolor}{rgb}{.8,.349,.1}
\usepackage{multirow}
\usepackage{color}
\usepackage[title]{appendix}%

\usepackage{float}
\usepackage[utf8]{inputenc}  
\usepackage{subfiles} 
\usepackage[square,numbers,sort&compress]{natbib}
\usepackage{amssymb}
\usepackage{mathtools}
\usepackage{gensymb}
\usepackage{authblk}
\usepackage[margin=0.8in]{geometry}
\usepackage[font=footnotesize]{subfig}
\usepackage{subfiles}
\usepackage{amsmath}
\usepackage[pdftex]{graphicx}

\usepackage{array}
\newcolumntype{M}[1]{>{\centering\arraybackslash}m{#1}}

\usepackage{hyperref}

\usepackage{multicol}
\usepackage{longtable}
\usepackage{forest}
\usepackage{adjustbox}
\usepackage{tikz}
\usetikzlibrary{arrows.meta, shapes.geometric, calc, shadows}
\usetikzlibrary{shapes,arrows,shadows}
\usetikzlibrary{shapes.arrows}

\colorlet{mygreen}{green!75!black}
\colorlet{col1in}{red!30}
\colorlet{col1out}{red!40}
\colorlet{col2in}{mygreen!40}
\colorlet{col2out}{mygreen!50}
\colorlet{col3in}{blue!30}
\colorlet{col3out}{blue!40}
\colorlet{col4in}{mygreen!20}
\colorlet{col4out}{mygreen!30}
\colorlet{col5in}{blue!10}
\colorlet{col5out}{blue!20}
\colorlet{col6in}{blue!20}
\colorlet{col6out}{blue!30}
\colorlet{col7out}{orange}
\colorlet{col7in}{orange!50}
\colorlet{col8out}{orange!40}
\colorlet{col8in}{orange!20}
\colorlet{linecol}{blue!60}

\pgfkeys{/forest,
  rect/.append style   = {rectangle, rounded corners = 2pt,
                         inner color = col6in, outer color = col6out},
  rectpy/.append style   = {rectangle, rounded corners = 2pt,
                         inner color = col2in, outer color = col2in},
  ellip/.append style  = {ellipse, inner color = col5in,
                          outer color = col5out},
  orect/.append style  = {rect, font = \sffamily\bfseries\LARGE,
                         text width = 325pt, text centered,
                         minimum height = 10pt, outer color = col7out,
                         inner color=col7in},
  oellip/.append style = {ellip, inner color = col8in, outer color = col8out,
                          font = \sffamily\bfseries\large, text centered}}
\usepackage{verbatim}
\usepackage{smartdiagram}
\usepackage{metalogo}
\usepackage{float}
\usepackage{listings}

\definecolor{backcolour}{rgb}{0.95,0.95,0.92}

\lstdefinestyle{mystyle}{
    backgroundcolor=\color{backcolour}, 
    breakatwhitespace=false,         
    breaklines=true,                 
    captionpos=b,                    
    keepspaces=true,                 
    showtabs=false,                  
    tabsize=2
}
\RequirePackage{subfig}
\RequirePackage{xspace}

\lstdefinelanguage{TelFortran}{                                                                          
  language=[90]Fortran,
  basicstyle=\ttfamily\small,
  frameround=ffff,
  stringstyle={\color{magenta}},
  showstringspaces=false,
  morekeywords={USE,PRECISION,INTENT,IN,OUT,INOUT,ONLY,ADVANCE,WHILE,POINTER,MODULE},
  deletekeywords=[1]{REAL},
  keywordstyle=\bfseries,
  commentstyle=\color{gray},
  morecomment=[l]{!},
  morekeywords=[2]{SQRT,REAL,INT,MAX,MIN},
  keywordstyle=[2]{\bfseries\color{violet}},
  morekeywords=[3]{ALLOCATED,PRESENT,DEALLOCATE,ALLOCATE},
  keywordstyle={\color{blue}},
  literate={%
    *{.EQ.}{{{\bfseries\color{green}.EQ.}}}{4}
     {.NE.}{{{\bfseries\color{green}.NE.}}}{4}
     {.LE.}{{{\bfseries\color{green}.LE.}}}{4}
     {.LT.}{{{\bfseries\color{green}.LT.}}}{4}
     {.GE.}{{{\bfseries\color{green}.GE.}}}{4}
     {.GT.}{{{\bfseries\color{green}.GT.}}}{4}
     {.NOT.}{{{\bfseries\color{green}.NOT.}}}{5}
     {.OR.}{{{\bfseries\color{green}.OR.}}}{4}
     {.AND.}{{{\bfseries\color{green}.AND.}}}{5}
     {.TRUE.}{{{\bfseries\color{violet}.TRUE.}}}{6}
     {.FALSE.}{{{\bfseries\color{violet}.FALSE.}}}{7}
  },
    escapeinside={(*@}{@*)},
}

\lstdefinelanguage{TelPython}{                                                                          
  language=Python,
  basicstyle=\ttfamily\small,
  frameround=ffff,
  stringstyle={\color{magenta}},
  showstringspaces=false,
  commentstyle=\color{gray},
  keywordstyle={\color{blue}},
}
\usepackage{tabularx}

\begin{document}

\title{Bayesian inference of numerical modeling-based morphodynamics: Application to a dam-break over a mobile bed experiment}%

\author[1,2]{Cédric Goeury}
\author{Fabien Souillé}

\affil[1]{\'{E}lectricité de France (EDF), Research and Development Division, National Laboratory for Hydraulics and Environment (LNHE), 6 Quai Watier, 78400 Chatou, France}

\affil[2]{Laboratoire d’Hydraulique Saint-Venant (LHSV), ENPC, Institut Polytechnique de Paris, EDF R\&D, 6 Quai Watier, Chatou, 78400, France}

\date{\today}
\maketitle

\begin{abstract}
Numerical modeling of morphodynamics presents significant challenges in engineering due to uncertainties arising from inaccurate inputs, model errors, and limited computing resources. Accurate results are essential for optimizing strategies and reducing costs. This paper presents a step-by-step Bayesian methodology to conduct an uncertainty analysis of $2$D numerical modeling-based morphodynamics, exemplified by a dam-break over a sand bed experiment. Initially, uncertainties from prior knowledge are propagated through the dynamical model using the Monte Carlo technique. This approach estimates the relative influence of each input parameter on results, identifying the most relevant parameters and observations for Bayesian inference and creating a numerical database for emulator construction. Given the computationally intensive simulations of Markov chain Monte Carlo (MCMC) sampling, a neural network emulator is used to approximate the complex $2$D numerical model efficiently. Subsequently, a Bayesian framework is employed to characterize input parameter uncertainty variability and produce probability-based predictions.
\end{abstract}

\section{Introduction}

\label{sec1}

Protection and management of surface water resources require an understanding of numerous potential impacts associated with sediment. Transport and morphodynamics can have a strong influence on hydraulic behavior (for instance decrease flood capacities of rivers \citep{Hou_2018} and impair navigation \citep{Orseau_2021}), ecological environment \citep{Zhu_2017}, among others. Thus, over the last decades, morphodynamic models have been increasingly used to model sediment transport in rivers \citep{Mezbache_2020}, coasts \citep{Van_Rijn_2003, Kazhyken_2024} and estuaries \citep{Pittaluga_2015}. In spite of the significant improvement in computational resources and the accuracy of numerical models, the need for realistic morphodynamic flow simulation is beyond the abilities of deterministic forecast \citep{Gomez_1999,Hu_2011} and the probabilistic approaches are of the utmost importance \citep{Goldstein_2019}. 

Uncertainty inherent to the numerical model typically arises from different sources such as (i) inaccuracies in the model inputs: model boundary conditions (hydrological inputs \citep{Oliveira_2021, Goulart_2023}, inflow sediment discharge \citep{Muller_2018, Beckers_2018}), description of morphodynamic parameters (sediment characteristics: critical Shields parameter \citep{Schmelter_2012,Mouradi_2016}, sediment grain size \citep{Pinto_2006, Villaret_2016}, and similar examples), initial conditions (geographical data: \citep{Siedersleben_2021}, bed composition \citep{Goulart_2023}, among others); (ii) errors in model structure, such as poorly described or omitted processes: empirical sediment transport formulae \citep{Pinto_2006, Goulart_2023}; and (iii) limited computing resources that constraints the number of parameters used to describe processes and the resolution of the model (spatial and temporal \citep{Fortunato_2009}).

Despite its importance, literature review shows only limited attention to uncertainty quantification with process-based morphodynamic models \citep{Villaret_2016,Vanderwegen_2013} and no concluding consensus regarding the results of these studies which are not generalizable from one case study to another \citep{Oliveira_2021}. Indeed, the outcomes of the uncertainty analysis are directly linked to the study's specifications and the characterization of the uncertain input parameters. To address this issue, a Bayesian methodology is carried out in this paper. Bayesian methods offer a theoretically solid general framework facilitating the quantification of parameter, structural and predictive uncertainty \citep{Wu_2009, Leisenring_2012, Schmelter_2012}. Initial uncertainties are represented as prior probability distributions, which are updated to posterior distributions using Bayes' rule after processing observations, thereby reducing uncertainty. However, Bayesian techniques based on sampling-based stochastic approaches, generally used due to their flexibility in handling complex models, suffer from computational burden. 

To overcome this challenge, integrating Bayesian inference with model reduction techniques can be helpful in finding practical solutions to the inverse problem of determining the causal factors from a given set of observations \citep{Mohammasi_2023}. Model reduction aims to provide a simplified representation of a complex model called emulator (or equivalently meta-model, reduced model response surface), fast to evaluate while being able to accurately approximate the underlying original complex model. In the field of sediment transport modeling with high computational time constraints, iterative Bayesian updating of a meta-model is generally adopted to limit the number of the full-complexity model as done in the framework of parameter estimation \citep{Beckers_2020, Mouris_2023} and model selection \citep{Mohammasi_2018}. In these previously mentioned papers, the meta-models build efficient probabilistic mapping from an input to an output space to only serve the purpose of accelerating model calibration and selection, not replace the full complexity model. 

The objective of this paper is to propose a comprehensive uncertainty analysis of bi-dimensional numerical modeling-based morphodynamics including a Bayesian estimate of uncertain parameter distribution. In this framework, the study is carried on a laboratory experiment corresponding to a  dam-break over a sand bed (bedload transport). The purpose of analyzing a laboratory case study is to minimize the overlapping morphodynamical effects that could obscure the analysis, while also leveraging the abundant available data. An originality of this work is the development of a simple three steps approach for establishing the Bayesian methodology. Step 1: the \textit{a priori} knowledge, which encompasses the information known to the researcher before data processing, is employed to identify influential morphodynamic model parameters and to select the most appropriate observations for the inference process. Step 2: with all of these elements, an artificial neural network (ANN) fast to evaluate is constructed to approximate with a good-quality the heavy computational full complex morphodynamic model. A neural networks is chosen in this study due to its flexibility and adaptability to various non linear cases (meteorology, oceanography, climate and weather, hydrology, etc.) \citep{Cherkassky_2006}. Step 3: the Bayesian framework is carried out to describe the morphodynamic parameter posterior probability distributions, allowing for the uncertainty propagation and the generation of probability-based predictions.

This paper is organized as follows. In Section \ref{sec:gov_eqs}, the governing equations for simulations are presented. Section \ref{sec:study_case} is focused to the problem specification with the application case introduction and its numerical description. Section \ref{sec:inv_prob} provides details of probabilistic materials used in this work. Section \ref{sec:results} describes the results. Section \ref{sec:discussions} is the discussion and Section \ref{sec:conclusions} is the conclusion.

\section{Governing equations}
\label{sec:gov_eqs}

This section introduces the mathematical model that governs the  hydro-morphodynamic evolution. The parameters of the model that can be considered uncertain are also introduced.

\subsection{Hydrodynamic formulation}
\label{subsec:hydro}

The Shallow Water Equations (SWE) are a two-dimensional depth-averaged version of the incompressible Navier-Stokes equations (vertical integration), with additional hypothesis about the flow such as negligible vertical scale compared to horizontal, negligible vertical acceleration, low bed slope, among others. The SWE are written as in Eq. \ref{eq:numerical:hydrodynamic:SaintVenant}, where $x$ and $y$ denote the Cartesian coordinates in the horizontal plane, and $t$ designates time.

\begin{equation}
\label{eq:numerical:hydrodynamic:SaintVenant}
\left\{
	\begin{array}{l}
	  \frac{\partial h}{\partial t} + \frac{\partial (hu)}{\partial x}  + \frac{\partial (hv)}{\partial y} = 0, \\
    \frac{\partial(hu)}{\partial t} + \frac{\partial(hu^2)}{\partial x} + \frac{\partial(huv)}{\partial y} = -gh\frac{\partial \eta}{\partial x} -  \frac{1}{\rho} {\tau_b}_x +  \nabla \cdot (h \nu_e \nabla u), \\
    \frac{\partial(hv)}{\partial t} + \frac{\partial(huv)}{\partial x} + \frac{\partial(hv^2)}{\partial y} = -gh\frac{\partial \eta}{\partial y} -  \frac{1}{\rho} {\tau_b}_y +  \nabla \cdot (h \nu_e \nabla v) .
	\end{array}
\right. 
\end{equation}

The system unknowns are the water depth $h(x,y,t) = \eta (x,y,t) - z_b(x,y,t)$, where $\eta$ and $z_b$ are the free-surface and bed elevations respectively, and $u$  and $v$ are the horizontal Cartesian components of the depth-averaged velocity $\mathbf{u}$ along the $x$ and $y$ axis respectively. The right hand side of the system contains a dispersion-diffusion term, assumed isotropic. Additionally, the transpose term $\nabla \mathbf{u}^T$ of the strain rate tensor is neglected.
The dispersion-diffusion is taken into account via a total effective viscosity $\nu_e = \nu + \nu_t + \nu_d $ which encompass kinematic viscosity ($\nu$), eddy viscosity ($\nu_t$), and \textit{dispersion} ($\nu_d$). Bed shear stress vector $\mathbf{\tau}_b$, with components ${\tau_b}_x$ and ${\tau_b}_y$ along the Cartesian coordinates $(x,y)$, corresponds to the bed resistance to the flow, which takes the form of near bed flow structures (small to big turbulent eddies), introducing a dissipative effect. The viscosity and bed shear stress plays a major role on the flow simulation \citep{Morvan_2008} and are hence capital to define for all environmental applications. Bed shear stress $\mathbf{\tau}_b$ is generally unknown, and is often written in the form of plane horizontal resistance to parallel flow, such that
\begin{equation}
\label{eq:bed_shear_stress}
    \mathbf{\tau}_b = \dfrac{1}{2} \rho C_f \mathbf{u} \sqrt{u^2+v^2},
\end{equation}
where $C_f$ is an non-dimensional friction coefficient determined using empirical estimates. In the current study, Manning's approximation written as in Eq. \ref{eq:stricklerEstimateOfShearStress} is used.
\begin{equation}
\label{eq:stricklerEstimateOfShearStress}
    C_f = \frac{2 g n^2}{h^{1/3}},
\end{equation}
where the Manning number $n$ is an empirical coefficient.

\subsection{Morphodynamic formulation}
\label{subsec:morphodyn}

Bottom elevation $z_b$ is present in SWE through the free surface variable $\eta$, and can evolve in time. This has a strong influence on flow modeling, and is of interest in morphodynamic problems, as for the studied case presented in this work. To describe morphodynamics, sediment particle motion is modeled, and distinction is usually made between \textit{suspension} that takes place in the water column, and \textit{bedload}, which is the rolling, sliding and hopping of particles near the bed. Both effects are accounted through a sediment mass balance equation named the Exner equation. This relationship results from mass conservation calculations on elementary bed surface \citep{Amoudry_2011}, where sediments are considered as a continuum. In this work, bedload is the main sediment transport process. When considering bedload without suspended sediment transport, the Exner equation can be written as presented in Eq. \ref{eq:manuscript:nearshore:numerical:morpho:exner}.

\begin{equation}
  \label{eq:manuscript:nearshore:numerical:morpho:exner}
  (1-\lambda) \frac{\partial z_b}{\partial t} + \nabla ~\cdot~\mathbf{Q}_b = 0,
\end{equation}

where bed elevation $z_b$ is transported with bed porosity $\lambda$ (or volumetric particle concentration $1-\lambda$), enhanced with volumetric bedload transport rate vector per unit width $\mathbf{Q}_b = \left(Q_{bx} = q_b \cos \left(\alpha\right) ,Q_{by} = q_b \sin \left(\alpha\right) \right)$ (m$^2$/s), $\alpha$ ($^{\circ}$) is the angle between the sediment transport vector and the downstream direction ($x$ axis) which is expressed as  $\alpha = \tan^{-1}\left(\frac{v}{u}\right)$ in absence of sediment transport corrections and $q_b$ is the bedload transport rate per unit width. In the following, considerations are done with scalar variables.

The transport rate $q_b$ can be calculated using several empirical formulas. In this study, the Meyer-Peter and Müller formula \citep{Meyer_1948} is used:
\begin{equation}
\frac{q_b}{\sqrt{g(s-1)d_{50}^3}}=\left\{\begin{array}{ll}
0 & \text{if}\,\theta<\theta_{cr}\\
\alpha_{MPM}(\theta-\theta_{cr})^{3/2} & \text{otherwise}
\end{array}
\right.
\label{eq:mpm}
\end{equation}
with $s = \frac{\rho_s}{\rho}$ the relative density ($-$); $\rho_s$ the sediment density (kg/m$^3$); $\rho$ the water density (kg/m$^3$); $d_{50}$ the median sand grain diameter and $g$ the gravity acceleration constant (m/s$^2$); $\alpha_{MPM}$ a transport coefficient; $\theta$ and $\theta_{cr}$ are respectively the Shields number and its critical value indicating the movement’s threshold. 

The Shields number $\theta$ is the dimensionless shear stress $\tau$ defined as $\theta = \frac{\tau}{g(\rho_s-\rho)d_{50}}$. The shear stress due to skin friction acting on bedload is expressed as follows:
\begin{equation}
  \label{eq:shear_stress}
  \tau = \frac{C'_f}{C_f} \sqrt{{\tau_b}_x^2 + {\tau_b}_y^2}
\end{equation}
where 
$C'_f=2\left(\frac{\kappa}{\ln \left(11.036\frac{h}{k_s}\right)}\right)^2$ (–) is 
the friction coefficient due to skin friction where $\kappa$ is the von Karman coefficient (-), $k_s=\alpha_{ks}d_{50}$ the roughness height (m), the coefficient $\alpha_{ks}$ is a calibration parameter (–).

The angle between the sediment transport vector and the downstream direction $\alpha$, as a consequence  of the downslope gravity force acting on the grains moving  along  a  sloping  bed, can be computed according to \citet{Struiksma_1985}:
\begin{equation}
  \label{eq:transport_angle}
  \tan \alpha = \frac{\sin \delta-\frac{1}{f(\theta)}\frac{\partial z_b}{\partial y}}{\cos \delta-\frac{1}{f(\theta)}\frac{\partial z_b}{\partial x}}
\end{equation}
with  the terms $\frac{\partial z_b}{\partial x}$ and $\frac{\partial z_b}{\partial y}$ representing respectively the transverse and longitudinal slopes, $\delta$ the angle between the sediment transport vector and the flow direction ($^{\circ}$) expressed as $\delta = \tan^{-1}\left(\frac{v}{u}\right)$ in absence of secondary currents and $f(\theta)$ is a function weighting the influence of the transverse bed slope, expressed according to \citet{Talmon_1995} as a function of the non-dimensional shear stress or Shields parameter as follow $f(\theta)=\beta_2 \sqrt{\theta}$, with $\beta_2$ is an empirical coefficient.

In addition to the transport direction, the streamwise bed slope effect must be taken into account in the magnitude of the sediment transport. In this study, the modification of the bedload transport rate by a factor that acts as a diffusion term in the bed evolution equation is considered as proposed by \citet{Koch_1980}:
\begin{equation}
  \label{eq:magnitude}
  q^{\star}_b = q_b\left[1-\beta\left(\frac{\partial z_b}{\partial x}\cos \alpha + \frac{\partial z_b}{\partial y}\sin \alpha\right)\right]
\end{equation}
where $q^{\star}_b$ is the modified bedload transport rate and $\beta$ is an empirical factor accounting for the streamwise bed slope effect.

\subsection{Model parameters}
\label{subsec:parameters}

To sum up, previous equations used empirical formulas that require the definition of scaling coefficients that are often unknown. This study is focusing only on morphodynamic parameters, including the porosity parameter ($\lambda$), the empirical sediment bedload transport formulae coefficients (the transport coefficient $\alpha_{MPM}$ and the sediment motion threshold $\theta_{cr}$), bedload magnitude and direction correction factors (proportional coefficient of the Nikuradse's equivalent sand-grain roughness $\alpha_{ks}$, transverse and streamwise bed slope effect coefficients, respectively, $\beta_2$ and $\beta$). The ranges of variation related to these parameters are provided hereafter.

\subsubsection{Porosity morphodynamic parameter}

The porosity term $\lambda$ in Eq. \ref{eq:manuscript:nearshore:numerical:morpho:exner} assesses the fraction of a given sediment volume that is composed of void space. In the current work, the material can be categorized as coarse sand in compliance with the sediment grade scale \citep{Garcia_2008}. Thus, the parameter variation interval is set to $[0.31, 0.46]$ \citep{Domenico_1998}.

\subsubsection{Sediment bedload transport formulae coefficients}

The Meyer-Peter and M\"uller bedload transport formula (Eq. \ref{eq:mpm}) is empirical based on uniform coarse sand and gravel laboratory experiments. However, it is commonly used for a wide range of sediment transport applications, also for cases which differ from its original design. For that purpose, the transport coefficient $\alpha_{MPM}$ is generally calibrated according to the application case. Based on literature review \citep{Wong_2006a,Won_2006b,Fernandez_Luque_1976,Meyer_1948,Wilson_1966,Zech_2008}, the factor $\alpha_{MPM}$ is assumed to be contained in an interval set to $[2.66,32]$.

Moreover, a threshold condition for sediment movement is existed and modeled by the critical Shields parameter ($\theta_{cr}$) in the bedload transport relationship (Eq. \ref{eq:mpm}). The establishment of justifiable values for critical Shields number is quite challenging, particularly when one accounts for the multiplicity of bed arrangements and varying hydraulic conditions that affect incipient motion \citep{Schmelter_2012}. In the current study, the sediment motion threshold is assumed to evolve within the interval $[0.022, 0.058]$ valid for spherical grains on Earth in low slope rivers \citep{Feehan_2023}. This range is supported by considering the extreme values of critical incipient motion ($[0.024, 0.057]$), which fall within the 10\% margin of the experimental median diameter $d_{50}$ and sediment density $\rho_s$, based on eight decades of incipient motion studies \citep{Buffington_1997}.

\subsubsection{Sediment bedload magnitude and direction correction factors}

The skin friction term is based on the concept of apparent roughness height (Eq. \ref{eq:shear_stress}). Since an uniform material is considered in this study, the Nikuradse's approach is well adapted to compute the  apparent roughness of the bed $k_s$. As stated by \citet{Garcia_2008}, the $\alpha_{ks}$ coefficient is taken to be within $[1,6.6]$.

The bedload transport direction correction is established on the basis of the transverse bed slope weighting function controlled by the parameter $\beta_2$ (Eq. \ref{eq:transport_angle}). This empirical factor is originally set to $0.85$ \citep{Talmon_1995}. However, this parameter has been calibrated in \citet{Yossef_2016} and \citet{Mendoza_2017} where optimal values of $0.2$ and $1.6$ were respectively found. Thus, the assessed interval for the transverse bed slope parameter is $[0.2,1.6]$.

The bedload transport magnitude correction is based on the formula proposed in \citet{Koch_1980}. From Eq. \ref{eq:manuscript:nearshore:numerical:morpho:exner} and \ref{eq:magnitude}, it can be noticed that the bed slope effect is similar to adding a diffusion term in the bed evolution equation. Several authors \citep{Parker_2003, Struiksma_1989} express $\beta$  as a function of the mean fluid bed shear stress. A higher $\beta$ coefficient results in greater diffusion of the solution. However, larger values can cause an unrealistic flattening of the profile. \citet{Walstra_2004} suggests a realistic range for the $\beta$ coefficient of $[0, 5]$. This range, which includes the one proposed by \citet{kopmann_2012}, is considered in this study.

\section{Presentation of the dam-break over a mobile bed scenario}
\label{sec:study_case}

This study examines an experiment involving dam-break flows over a sand bed. This experimental case serves as a robust and well-documented basis for uncertainty quantification, offering a dynamic and complex environment to validate predictive capabilities and enhance model accuracy. This section outlines the details of the experiment and the numerical modeling of dam-break dynamics.

\subsection{Experiment configuration}
\label{subsec:exp}

\begin{figure}[h]
\begin{center}
\includegraphics[width=0.85\textwidth]{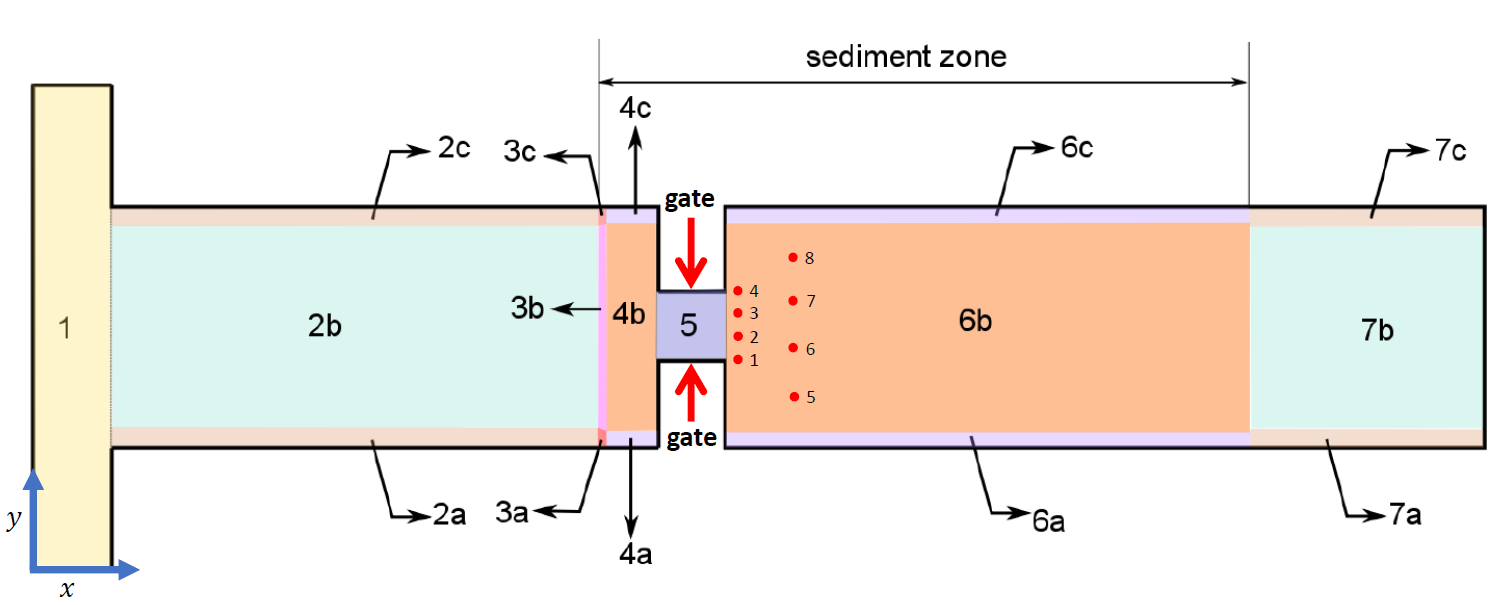}
\caption{Flume experiment configuration for the two dimensional dam-break with measurement probe locations ({\color{red}$\bullet$}) \citep{Soares_2012}}
\label{fig:dam-break_flume}
\end{center}
\end{figure}

Experiments of dam-break flows over a sand bed were conducted at Université catholique de Louvain, Belgium \citep{Soares_2012}. From these experiments, a benchmark was proposed to the scientific community. The set-up of the experiment is stretched in Figure \ref{fig:dam-break_flume}. The whole useful length of the flume is around $27.59$ meters, with a width of $3.6$ meters. The original axes $(x,y)$ are taken as $(0, 0)$. The sub-zones with their distinct bathymetry characteristics are provided in Appendix \ref{secA1:flume_bathymetry}. The dam is considered at $x = 12.09$ m. The $z$ elevation is taken with reference to the fixed bed (areas 1, 2, 3 and 7, $z = 0$ m). The sediment zone is located at areas 4 to 6, with a sediment layer thickness of $0.085$ m, which extends $1.5$ m to the upstream of the dam and $9$ m downstream. A gradual transition sediment layer thickness is made over $20$ cm at the beginning of the zone with sediment number $4$. The size of uniform sand used in the experiment is characterized by $d_{50} = 1.61$ mm, with its relative specific gravity to water of $2.63$ and initial porosity of $0.42$. 

To simulate the dam-break, the gate located at $12$ m from the upstream flume end was pulled up rapidly to reproduce an instantaneous release. The water level upstream ($x<12.09$) is set to $0.47$ m, $0.085$ on the sand layer which corresponds to an initially saturated sand layer and $0$ otherwise. As observed in the experimentation, bedload transport, involving sediments moving along the bed without losing contact, is the main sediment transport process \citep{Soares_2012}.

\begin{figure*}[!h]
\centering
\includegraphics[width=0.7\textwidth, trim = 1cm 1cm 1cm 0.5cm]{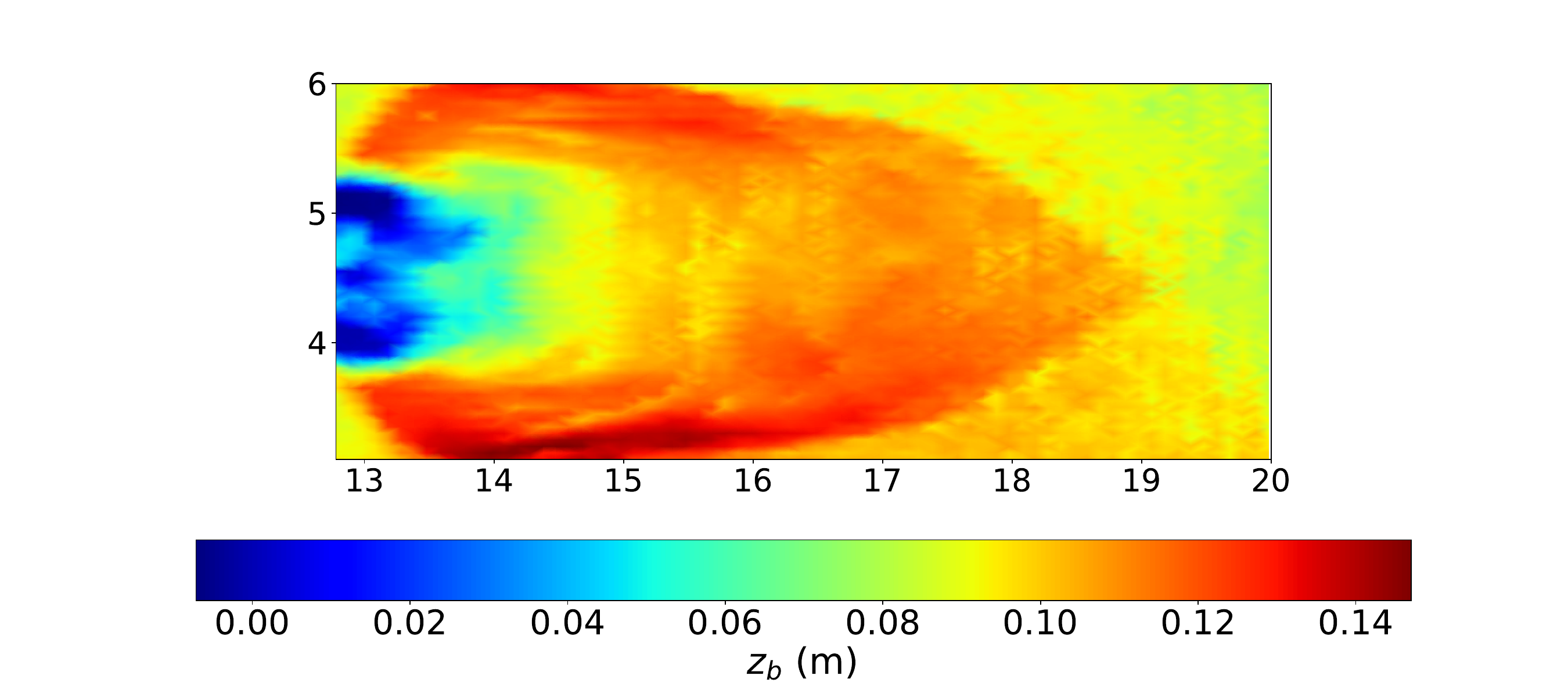}
\caption{Final bed topography reconstructed from the measured bed proﬁles}
\label{fig:raw_zf_exp}
\end{figure*}

The boundary conditions consist at a closed wall upstream and a free outflow downstream of the flume. At time $t=0$, the gate is rapidly lifted, The passage of the dam-break flow generates important morphological changes. After 20 s of flow, the gate is closed in order to stop the experiment. The temporal water level evolution was measured using 8 ultrasonic probes (Figure \ref{fig:dam-break_flume}). In addition, the final bed topography was measured using a bed profiler every $\Delta y = 0.05$ m (Figure \ref{fig:raw_zf_exp}).

\newpage

\subsection{Numerical configuration}
\label{subsec:numerical}

In this work, coupled simulations were used aligning the hydrodynamic 2D model, TELEMAC-2D \citep{Hervouet_2007}, and the sediment transport and bed evolution model GAIA \citep{Tassi_2023}, which are part of the openTELEMAC numerical platform (\url{www.opentelemac.org}). These module are used to solve respectively the Shallow Water (Eq. \ref{eq:numerical:hydrodynamic:SaintVenant}) and Exner (Eq. \ref{eq:manuscript:nearshore:numerical:morpho:exner}) equations complemented by boundary conditions. They relies on a continuous, piecewise-linear finite element approach, based on an unstructured triangular mesh. The N edge-by-edge scheme and the Positive Streamwise Invariant distributive scheme are used to compute the advection of velocity and water depth, respectively. The Exner equation fluxes are also computed with a N edge-by-edge distributive scheme with a limiter to ensure the positivity of the erodible layer.

The numerical model is composed of $24 536$ triangular finite elements corresponding to $0.1$ m of spatial discretization. The time step is set to $0.01$ s providing a Current number (defined from the Shallow Water equations) less than $0.2$ in the simulation. The Manning coefficient is set to 0.0165 s/m$^{1/3}$ corresponding to the value experimentally evaluated under uniform flow conditions \citep{Soares_2012}. For the 2D numerical modeling, the upstream boundary condition is a closed wall. A Neumann-type boundary condition (i.e. free height and velocity) is imposed on the downstream end. The sediment transport is assumed to be induced by bedload. For this purpose, the Meyer-Peter and M\"uller formula is considered in the following statements. To summarize, the following physical and numerical parameters are presented in Appendix \ref{secA2:num_and_phys_config}.

\section{Inverse problem description}
\label{sec:inv_prob}

The inverse problem can be understood as the computation of the \textit{a posteriori} distribution $\pi\left(\xi|Y^{obs}\right)$ providing the range within which a parameter is likely to fall, based on observed data and prior information, reflecting the uncertainty in its estimation. The \textit{posterior} distribution can be determined through the well-known Bayes rule (Eq. \ref{eq:bayes}).

 \begin{equation}
 \pi\left(\xi|Y^{obs}\right)=\frac{\pi\left(Y^{obs}|\xi\right)\pi\left(\xi\right)}{\int\pi\left(Y^{obs}|\xi\right)\pi\left(\xi\right)d\xi}
 \label{eq:bayes}
 \end{equation}

The term $\pi\left(Y^{obs}|\xi\right)$, called the likelihood, can be interpreted as a measure of the information provided by the observations $Y^{obs}$ defined on a probability space $\left(\Omega,B,\mathbb{P}\right)$ (with $\Omega$ as a sample space, $B$ the $\sigma$-algebra of events and $\mathbb{P}$ the probability measure), conditional upon the p-components of the parameter control vector $\xi\in\mathbb{R}^p$ composed of variables also defined on a probability space. Around the unknown parameter vector $\xi$ and the observational space $\mathbb{R}^m$ is the observation space defined by Eq. \ref{eq:observation_space}.

 \begin{equation}
     Y^{obs}=M\left(\zeta\right)+\epsilon_o
     \label{eq:observation_space}
 \end{equation}

 where $M:\mathbb{R}^d\longrightarrow\mathbb{R}^m$ is a vector-valued function of vector $\zeta$ and $\epsilon_{obs}\in\mathbb{R}^m$ is an assumed observable additive measurement noise such as $\mathbb{E}\left(\epsilon_{obs}\right)=0$ and $R =cov(\epsilon_{obs})=\mathbb{E}\left(\epsilon_{obs}\epsilon_{obs}'\right)\in\mathbb{R}^{m\times m}$ and identified as a multivariate normal distribution, $\epsilon_{obs} \sim \mathcal{N}\left(0,R\right)$. In this work, for simplicity an unbiased additive Gaussian error model is considered, thus, the error observation covariance matrix $R$ is taken diagonal such as $R = \sigma_oI_{N_o}$.

The term $\pi\left(\xi\right)$ represents \textit{a priori} knowledge of the unknown parameters $\xi$ considered independent hereafter. In this work, the standard deviation $\sigma_o$ is considered as a parameter to infer since it includes imprecise measurement and representativity errors. A half-normal distribution, chosen for its non-negativity, represents prior knowledge of $\sigma_o$, reflecting the belief that measurement noise is usually small but can sometimes be larger. Finally, the unknown parameter vector $\xi$ is defined such as $\xi=\left(\zeta,\sigma_o\right)$.

The integral in the denominator of the Eq. \ref{eq:bayes} makes it impossible to get a closed-form expression for the posterior distribution $\pi\left(\xi|Y^{obs}\right)$. To overcome this issue, Markov chain Monte Carlo (MCMC) method can be used to get a sample from that distribution or a distribution that approximates it reasonably well. Then, inference and estimates from it can be carried out. Markov chain Monte Carlo (MCMC) algorithms may require long time to converge to the posterior distribution since they are based on a random walk process in parameter space exploration. Hamiltonian Monte Carlo (HMC), used in this work, is a Markov chain Monte Carlo (MCMC) algorithm that mitigates this issue by incorporating first-order gradient information into the sampling process, thereby avoiding the random walk behavior.

\subsection{Hamiltonian Monte Carlo method}
\label{subsec:HMC}

The HMC method, as its name suggests, is directly derived from Hamiltonian mechanics which is a reformulation of Lagrangian mechanics as described in Appendix \ref{sec:uq:subsub:HM}. Appealing to the physical analogy, the canonical distribution from statistical mechanics is used to described the Hamiltonian, an energy function, in terms of probabilities as defined in Eq. \ref{eq:defcano}.

\begin{equation}
\pi\left(\xi,\textbf{p}\right) = A\exp{\left(-\mathcal{H}\left(\xi,\textbf{p}\right)\right)}
\label{eq:defcano}
\end{equation}

with $A$ is the normalizing constant and $\textbf{p}$, also called the momentum variables, is introduced to expand the $D-$dimensional parameter space $\xi$ into a $2D-$dimensional phase space $(\xi, \textbf{p})$. 

The Hamiltonian is formulated as $\mathcal{H}\left(\xi,\textbf{p}\right)=K(\xi,\textbf{p})+V(\xi)$ where $K(\xi,\textbf{p})$ and $V(\xi)$ represent kinetic and potential energy, respectively. The potential energy $V(\xi)$ is expressed from the posterior distribution such as $V(\xi)=-\log\left(\pi\left(\xi|Y^{obs}\right)\right)$ simplified to $V(\xi)=-\log\left(\pi\left(\xi\right)\right)$ hereafter for the sake of clarity. The kinetic energy, defined as $-\log\left(\pi\left(\textbf{p}|\xi\right)\right)$ based on the joint probability $\pi\left(\xi,\textbf{p}\right)$ and potential energy $V(\xi)$, is usually modeled using a Gaussian distribution over the momentum variables (Eq. \ref{eq:kin_choice}), as this approach empirically performs better than other choice \citep{Betancourt_2018}.

\begin{equation}
\pi\left(\textbf{p}|\xi\right) = \mathcal{N}\left(\textbf{p}|0,M\right)
\label{eq:kin_choice}
\end{equation}

Due to volume conservation property in the parameter-momentum phase space $(\xi, \textbf{p})$ \citep{Neal_2012}, the choice of the momentum Gaussian distribution's covariance matrix $M$  affects the parameters $\xi$. To facilitate exploration of Hamiltonian level sets, the inverse of momentum covariance matrix $M$ is set to the target distribution's covariance, estimated during the MCMC warm-up stage.

From all ingredients previously defined, the Hamilton's equations (Eq. \ref{eq:HM_equations}), can be numerically integrating in time with the Leapfrog integrator for instance. Due to numerical inaccuracies during time integration, the Hamiltonian trajectory is used as a proposal function for a Metropolis-Hastings algorithm \citep{Hastings_1970}. After negating the momentum variables to ensure reversibility, an acceptance-rejection step is performed to converge to the target distribution. Each proposed state $(\xi^\star, \textbf{p}^\star)$ is accepted with probability $\min\{1,\exp{[-\mathcal{H}\left(\xi^\star, \textbf{p}^\star\right)+\mathcal{H}\left(\xi, \textbf{p}\right)]}\}$. If rejected, the current state $\left(\xi, \textbf{p}\right)$ remains unchanged. Finally, the target distribution is derived from HMC sampling by projecting away the momentum, leveraging the independence of the joint distribution. 

To avoid the complexity of tuning numerical integration, this work uses the No-U-Turn Sampler (NUTS) extension of HMC \citep{Hoffman_2014}, which eliminates the need for hyper-parameter tuning. The ergodicity and posterior stationarity of the Markov chain produced by HMC and NUTS are demonstrated in \citet{Durmus_2023}.

\subsection{Practical strategies for physical link implementation}
\label{sub:practial_part}

To perform the MCMC inversion, the likelihood and prior distributions must be specified \textit{a priori} (Eq. \ref{eq:bayes}). 

As presented in Section \ref{sec:gov_eqs}, morphodynamic models depend on empirical sediment transport predictors from small-scale experiments, requiring appropriate parameter selection for each study case. In this context, the parameter control vector $\zeta$, emerging in Eq. \ref{eq:observation_space}, is composed of the morphodynamic parameters described in Subsection \ref{subsec:parameters}. The model parameter uncertainties are described by uniform distributions, based on the principle of maximum entropy, defined by each parameter's minimum and maximum values. Moreover, as TELEMAC-2D/GAIA is a spatially and time discrete model, for a realization $\zeta^j$, the model output $Y$ in agreement with the experiment measurements (Subsection \ref{subsec:exp}) is either composed of topography scalar output given on a set of $m\in\mathbb{N}$ space coordinates $\left(x_1,...,x_m\right)$ at the final simulation time $T$ such as $Y=\left(Y_1,...,Y_m\right)$ or, for each point of interest corresponding to probe locations (${\bf{x}}_{p}$), the water level scalar output given at discrete time $t\in[1,…,T]$ such as $Y=\left(Y_1,...,Y_T\right)$ (with respectively $Y_i=Y({\bf{x}}_i,T)= M(\zeta^j;\left({\bf{x}}_i,T\right))$ and  $Y_i=Y({\bf{x}}_{p},t_i)= M(\zeta^j;\left({\bf{x}}_{p},t_i\right))$).

At this stage, relevant key questions include the effects of parameters ($\zeta$) on simulated state variables compared to observations ($Y$), and which observations ($Y^{obs}$) to include in the inference process. In this work, these questions are addressed by a global sensitivity analysis presented in Subsection \ref{subsec:sensitivity}.

From all elements determined from the sensitivity analysis, the Bayesian method can be applied. However, in real-world applications, Markov chain Monte Carlo (MCMC) sampling-based approaches are computationally intensive and time-consuming. Still, the chosen HMC method requires gradient of the log-posterior involving computing the adjoint of the model TELEMAC-2D/GAIA operator. The partial derivatives of the complex model with respect to its input parameters are at best tedious and at worst impossible to compute.

To overcome these issues, a simplified representation of a complex model called emulator, fast to evaluate while being able to accurately approximate the underlying original complex model, is used in the current study. As detailed in Subsection \ref{subsec:rom}, a standard architecture Neural Network (NN) is carried out based on an open-source machine-learning library PyTorch \citep{Paszke_2019}. In addition to the benefits of this open-source library, PyTorch offers a large number of interoperable libraries which simplify the development of specific machine learning applications. Among them, the Deep universal probabilistic programming with Python and PyTorch library Pyro \citep{Bingham_2019} is used in this work for the Hamiltonian Monte Carlo method. Since Pyro is based on PyTorch, it benefits from tensor library with algorithm differentiation for gradient estimation.

\subsubsection{Sensitivity analysis}
\label{subsec:sensitivity}

The sensitivity analysis aims at quantifying the impact of uncertainty in input variables on the accuracy of the model output variables. The non-linearity between processes in morphological models may affect only a small proportion of input parameter space and leads to output variability that is difficult to translate by any moment of the random variable. As a consequence, sensitivity analysis must satisfy the following requirements \citep{Saltelli_2002,Borgonovo_2007}: global (i.e. entire input distribution taken into consideration), model free (i.e. no assumptions on the model functional relationship to its inputs), moment independence (i.e. avoid some information loss by the use of an output variability summary) and be quantitative. For the sake of clarity, in the following, the method is described in the case of scalar output such as $Y_p=Y({\bf{x}}_p,t_k)= M(\zeta_1,\zeta_2,...,\zeta_d;\left({\bf{x}}_p,t_k\right))$ where ${\bf{x}}_p$ is the location of the variable of interest extracted at the time step $t_k$. In this framework, moment-independent sensitivity measures based on the shift between the output distribution and the same distribution conditionally to a parameter was proposed by \citet{Borgonovo_2007} and read as:

\begin{equation}
    \delta_i = \frac{1}{2}\left[\int\pi_{Z_i}\left(\zeta_i\right)\left[\int\big|\pi_Y\left(y\right)-\pi_{Y|Z_i}\left(y\right)\big|dy\right]d\zeta_i\right]
\end{equation}

A Borgonovo sensitivity index value close to 0 and, conversely, 1 means that the output is, respectively, independent or highly dependent on the variable of interest. Based on the entire distribution, the Borgonovo sensitivity analysis can become unsuitable when the computational cost of each run of the model is non negligible or the number of model inputs is large. Thus, a sample-based estimation, performed in this study with a Monte Carlo sampling, was proposed to handle this issue \citep{Plischke_2013}. In the present work, the analysis was carried out based on an open-source library for performing sensitivity analyses ``SALib'' \citep{Herman_2017}.

\subsubsection{Reduced Order Modeling via Neural Networks}
\label{subsec:rom}

Neural Networks (NN) are widely used in environmental applications like hydraulic \citep{Pianforini_2024} and sediment transport \citep{Samantaray_2018} due to their strong modeling and prediction abilities. Their adaptable architecture makes them effective for complex, non-linear problems \citep{Sengupta_2020}. This work uses a standard neural network (NN) architecture corresponding of many connected units producing a sequence of real-valued activations coming from a non-linear function apply to the unit's input weighted connection. The first input layer contains the set of inputs $\zeta$, transformed through the second layer (first hidden layer) via the activation function, and progressively communicated through the network until the information reaches the last output layer to constitute the interest outputs $Y$.

To ensure the accuracy and generalizability of the neural network emulator, the high fidelity simulation results, coming from a Monte Carlo sampling of the TELEMAC-2D/GAIA model, are partitioned into distinct subsets for training, validation and testing.

The goal of the training step is to minimize the difference between the predicted output from the neural network $\tilde{M}$ and the desired output $M$. In the present study, the Adam stochastic optimizer is chosen to adjust the  weights and biases of the network from a mean square error loss function. The optimizer algorithm works iteratively by randomly select an ensemble of data point from the training set called a batch. It computes the gradient of the loss function with respect to the selected data and adjusts the parameters in the opposite direction of the gradient. The process is repeated until the entire training data set is exhausted completing one training epoch. The training is performed for a sufficient number of epochs to obtain a converged network. The convergence speed of the training is controlled by a learning rate. 

The architecture design of Neural Network is a challenging task as the number of neurons is known to compensate for the number of layers (and vice-versa). Indeed, the question of narrow (number of neurons per layer) vs. shallow (number of layers) networks remains open \citep{Lu_2017}. Thus, to get an accurate memory model, the topology of the neural network is determined based on the tree-structured Parzen estimator (TPE) optimization process from the open source hyperparameter optimization framework to automate hyperparameter search library ``Optuna'' \citep{Akiba_2019}. Thus, in the current work, the number of layers and the number of hidden units at each layer in the neural network is chosen to minimize the validation error. As its name suggests, this error is estimated from the validation data set of the trained networks. 

At the end of the previous stage, a neural network is constructed and designed respectively based on train and validation subsets. Its performance is then evaluated on the unseen test subset with the computation of the spatial predictivity criterion defined by the equation \ref{eq:Q2_spatial}.

\begin{equation}
  Q^2\left({\bf{x}}\right)=1-\frac{\sum_{j=1}^{n_{test}}\left[\hat{M}\left(\zeta_j,{\bf{x}}\right)-M\left(\zeta_j,{\bf{x}}\right)\right]^2}{\sum_{j=1}^{n_{test}}\left[\hat{M}\left(\zeta_j,{\bf{x}}\right)-\frac{1}{n_{test}}\sum_{i=1}^{n_{test}}M\left(\zeta_i,{\bf{x}}\right)\right]^2}
  \label{eq:Q2_spatial}
\end{equation}

\section{Application of the Bayesian framework}
\label{sec:results}

\begin{figure*}[!h]
\centering
\includegraphics[trim = 2.5cm 0cm 2.5cm 0cm, width=0.95\textwidth]{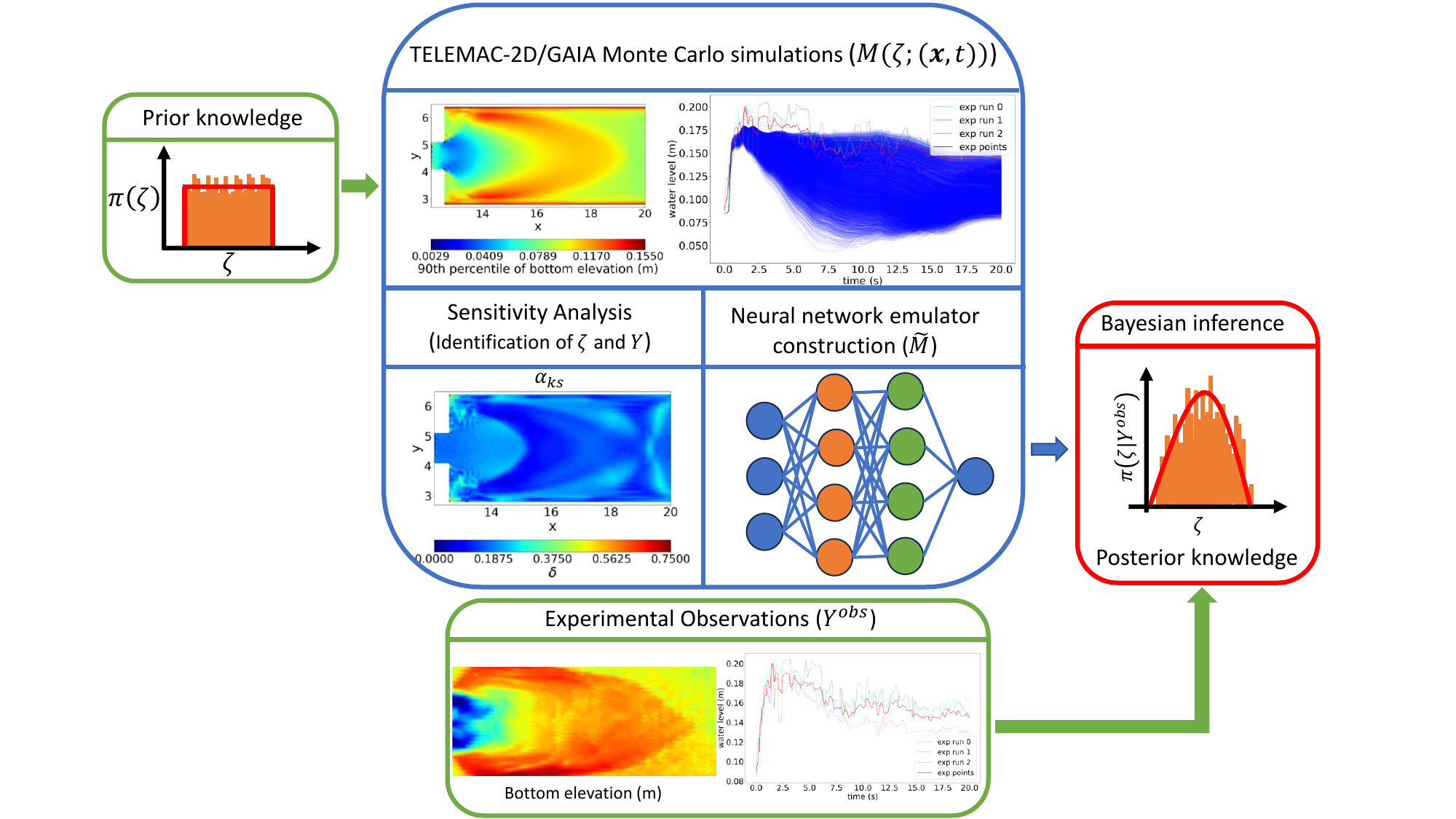}
\caption{Bayesian inference workflow}
\label{fig:summary}
\end{figure*}

In summary of the methodology previously described and illustrated in the Figure \ref{fig:summary}, prior knowledge is initially used to (i) identify influential morphodynamic model parameters, (ii) select the most appropriate observations for the inference process (Section \ref{res:sub:sensitivity}) and (iii) construct an efficient artificial neural network (NN) to approximate with a good-quality the heavy computational full complex morphodynamic model (Section \ref{subsub:res:NN}). For that purpose, the Monte Carlo technique is used to propagate input uncertainties through the model into the output quantity of interest as presented in \citet{Goeury_2022}.

Then, in Section \ref{sub:res:bayes_infer}, the Bayesian sampling method is applied to a twin experiment to evaluate the approach’s performance, and subsequently to the flume measurement-based experiments.

\newpage

\subsection{Prior knowledge exploitation}
\label{res:sec:prior}

\subsubsection{Sensitivity analysis}
\label{res:sub:sensitivity}

\begin{figure}[h]
    \centering
    \includegraphics[width=0.95\textwidth]{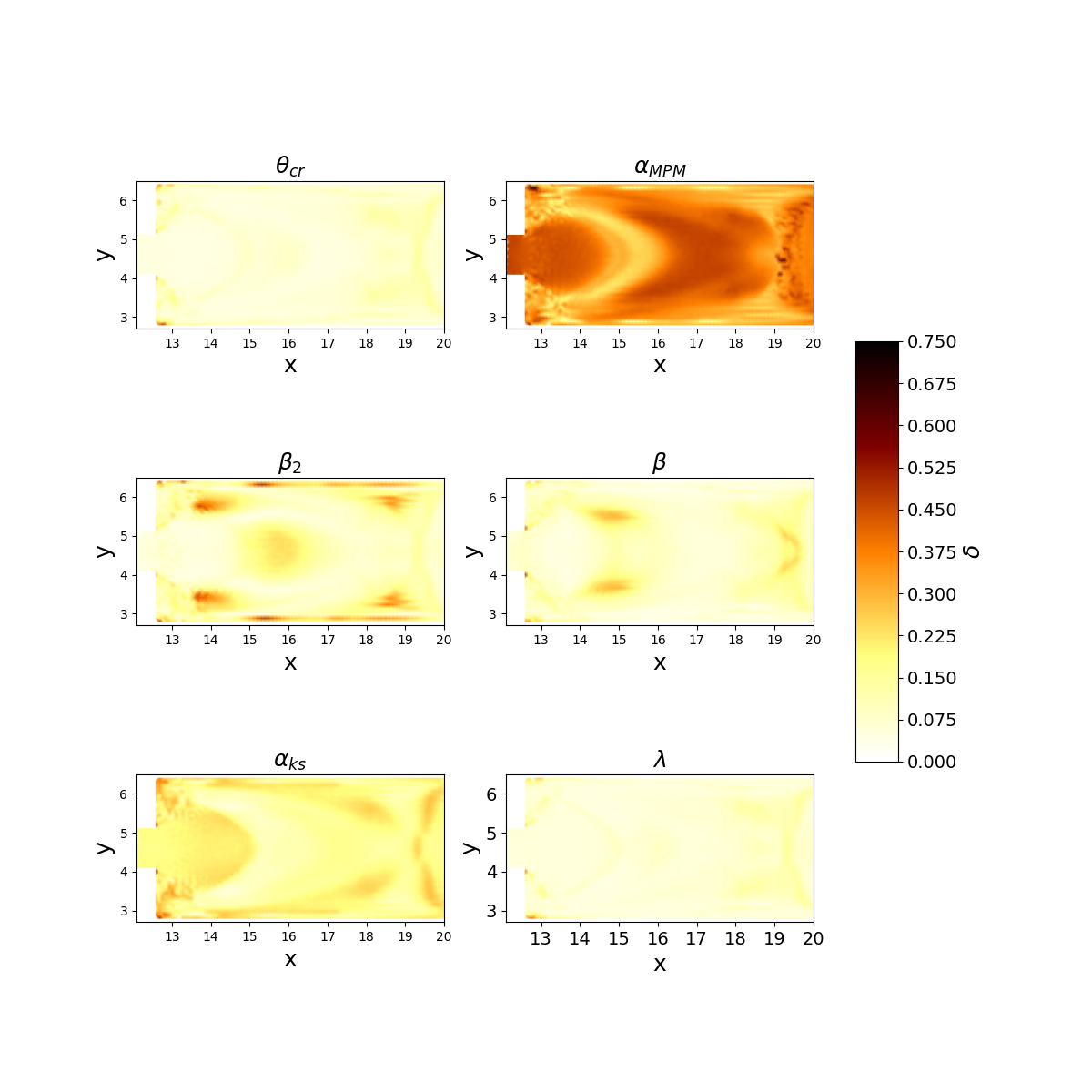}
    \caption{Borgonovo sensitivity indices in the computation domain for each uncertain variable.}
    \label{fig:borgo_all_variables}
\end{figure}

As mentioned in Section \ref{subsec:sensitivity}, sensitivity indices are approximated using a Monte Carlo sample. A convergence analysis, presented in Appendix \ref{sec:appendix:as}, confirmed that $7000$ model evaluations are needed for accurate and robust sensitivity estimates. Figure \ref{fig:borgo_all_variables} displays how each input variable affects bottom elevation changes across all computational domains. \newpage

\begin{figure*}[!h]
\centering
\begin{tabular}{cc}
\subfloat[Time evolution at P1 location\label{fig:time_series_us1}]{%
\includegraphics[width=0.45\textwidth]{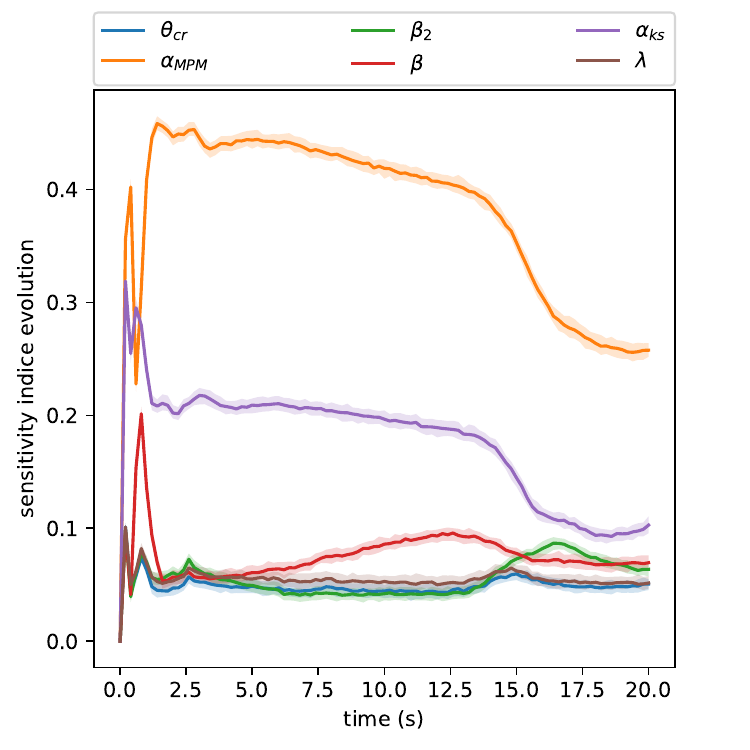}}&
\subfloat[Spatial evolution at the cross-section $y = 5.3$ m ($+70$ cm from the middle longitudinal axe) at the final experiment time\label{fig:cross-section}]{%
\includegraphics[width=0.45\textwidth]{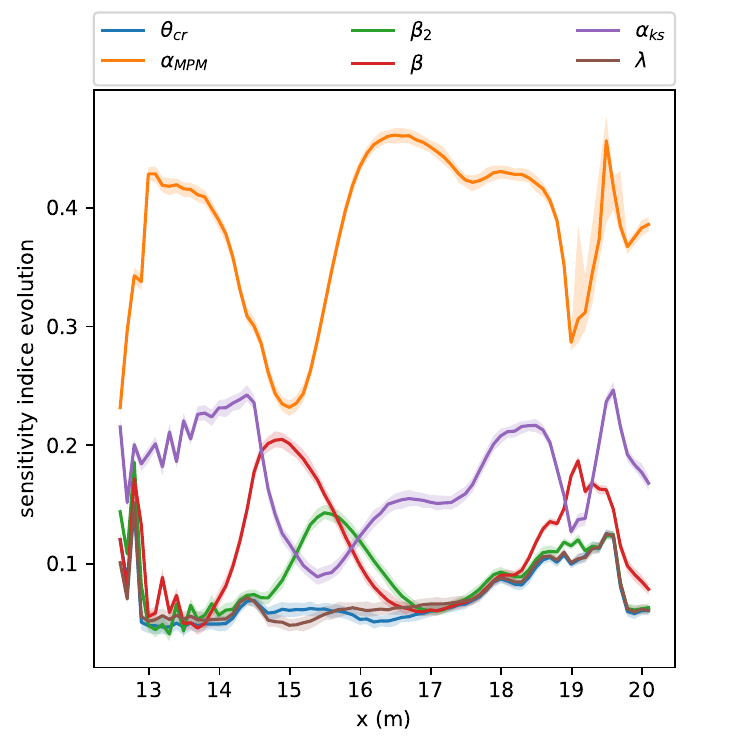}}\\
\end{tabular}
\caption{Spatial and time evolution of Sensitivity analysis indices:  (a) Time evolution of sensitivity indices at probe position P1 for the water elevation output, (b) Spatial evolution of sensitivity indices along a $y$ axis cross section for bottom elevation output}
\label{fig:AS_point_and_section}
\end{figure*}

Figure \ref{fig:borgo_all_variables} shows that bottom elevation is primarily influenced by the sediment bedload transport coefficient ($\alpha_{MPM}$) and, to a lesser extent, by the sediment bedload magnitude and direction correction factors ($\alpha_{ks}$, $\beta_2$, and $\beta$). The contributions of porosity ($\lambda$) and critical shear stress ($\theta_{cr}$) are negligible and thus excluded from further analysis. These results can be physically interpreted as follows. During the dam-break experiment, high flow velocities and shear stresses exceed the critical shear stress threshold, making sediment transport highly dependent on the bedload transport coefficient ($\alpha_{MPM}$). The spatial distribution of sensitivity indices shows that $\alpha_{MPM}$ and $\alpha_{ks}$ significantly influence areas of erosion and deposition, while $\beta_2$ and $\beta$ mainly affect the morphodynamic at the edges of bed forms. Figure \ref{fig:cross-section} further illustrates that the sensitivity indices for $\alpha_{MPM}$ and $\alpha_{ks}$ exhibit similar evolution structures, which are inversely related to the structures of $\beta_2$ and $\beta$. In complement to the morphodynamic parameter analysis, Figure \ref{fig:AS_point_and_section} presents the sensitivity analysis for the free surface and bottom elevations, measured during the experiment.

\newpage

Figures \ref{fig:time_series_us1}$-$\ref{fig:cross-section} show that spatial sensitivity analysis based on bottom elevation captures more variability in morphodynamic parameters than temporal analysis based on the free surface. This indicates that the phenomena exhibit stronger spatial variability, making spatial bottom elevation data more suitable for inverse problem characterization.

 In summary, the spatial bed elevation measurements at the end of the experiment are used to identify the most influential morphodynamic parameters $\zeta=\left(\alpha_{MPM}, \alpha_{ks}, \beta_2, \beta\right)$ using the previously described Bayesian process.

 \subsubsection{Neural network emulator}
\label{subsub:res:NN}

The neural network built for this study has a multi-layer architecture, designed to directly learn the bathymetry fields ($\textbf{z}_b$) at the end of the experiment from a set of parameters ($\zeta$). Based on the sensitivity analysis convergence study (Appendix \ref{sec:appendix:as}), $7200$ high-fidelity simulation results were selected for the training dataset. The sizes of the validation and test datasets were arbitrarily set to represent $25$\% and $12.5$\% of the training one, respectively. Input and output features are scaled such as the minimum of feature is made equal to zero and the maximum of feature equal to one. The current study uses $5000$ epochs, with training data shuffled and divided into mini-batches of $64$, and a learning rate of $1\times 10^{-4}$. The neural network topology is searched within $1$ to $10$ hidden layers and $4$ to $128$ units per layer. All layers, except the last, use a ``relu'' activation function. After $1200$ trials of TPE optimization process, the optimal neural network structure is made up of $5$ hidden layers composed respectively with $113$, $88$, $116$, $128$ and $124$ hidden units.

\begin{figure}[!h]
\centering
\begin{tabular}{cc}
\subfloat[TELEMAC-2D/GAIA result\label{fig:twin_exp_telemac}]{%
\includegraphics[trim = 1.5cm 1.5cm 1.cm 4.5cm, clip, width=0.35\textwidth]{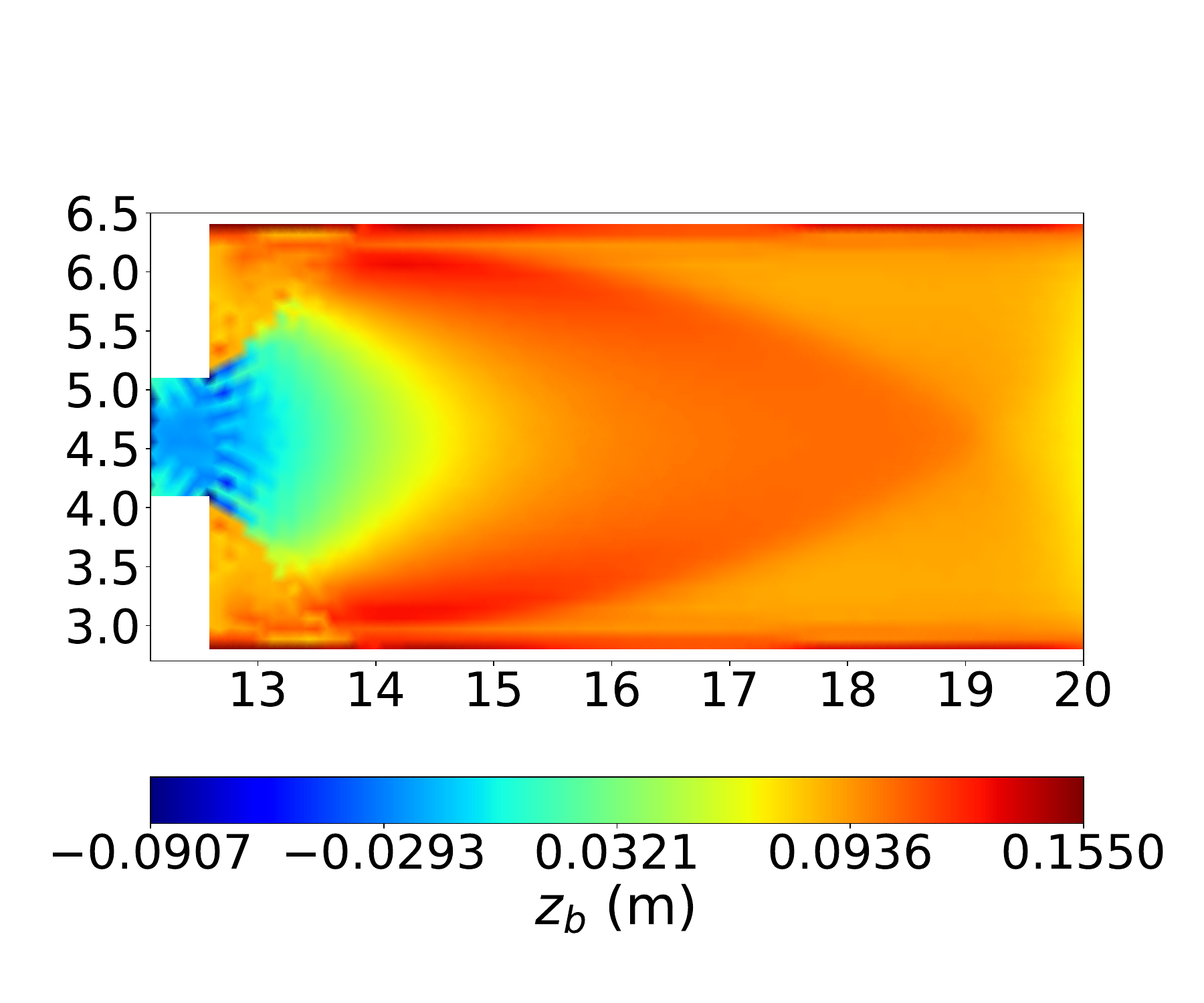}}&
\subfloat[Neural Network result \label{fig:twin_exp_nn}]{%
\includegraphics[trim = 1.5cm 1.5cm 1.cm 4.5cm, clip, width=0.35\textwidth]{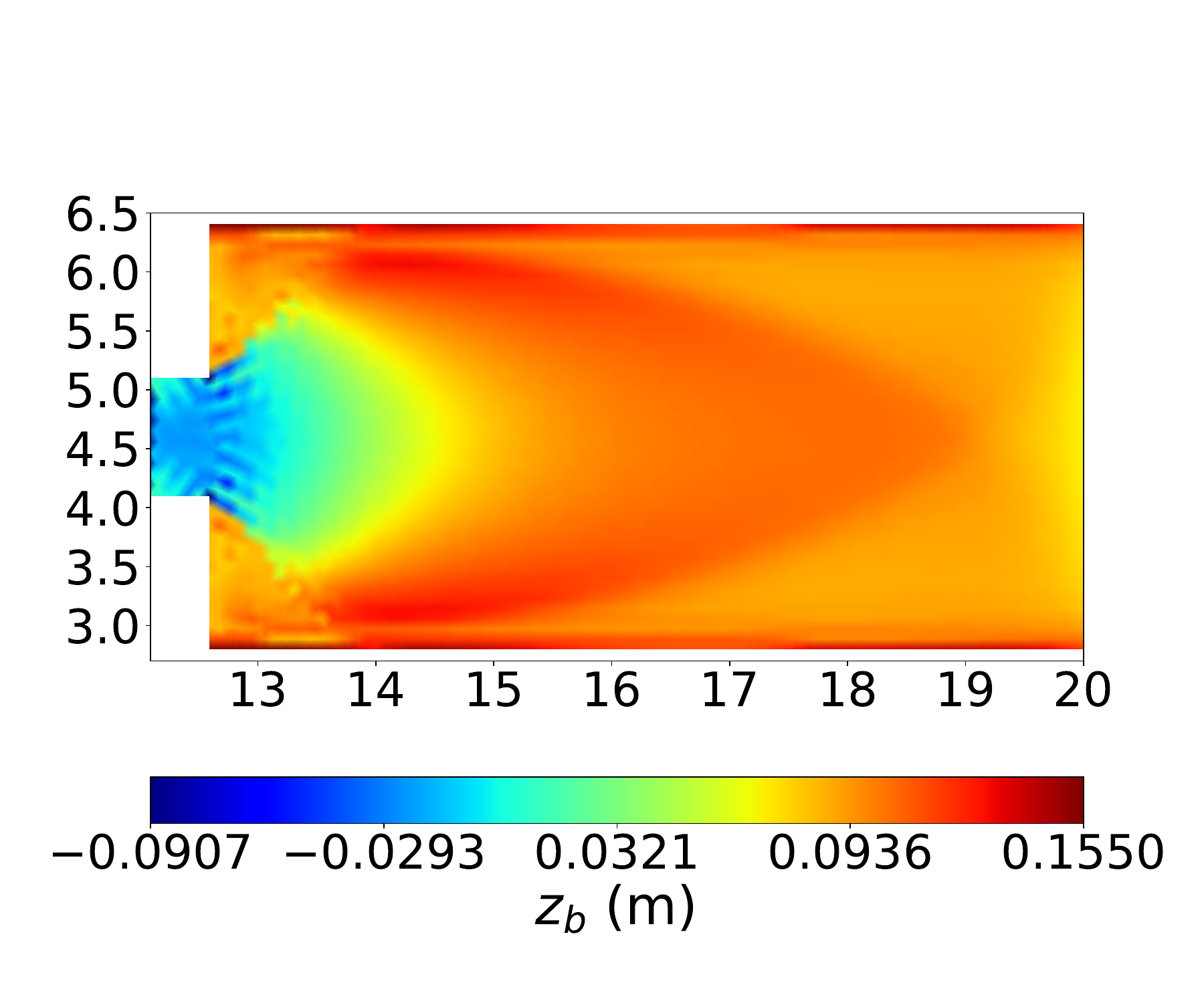}}\\
\end{tabular}
\subfloat[Predictivity criteria $Q^2$ \label{fig:predictivity}]{%
\includegraphics[trim = 1.5cm 1.5cm 1.cm 4.5cm, clip, width=0.35\textwidth]{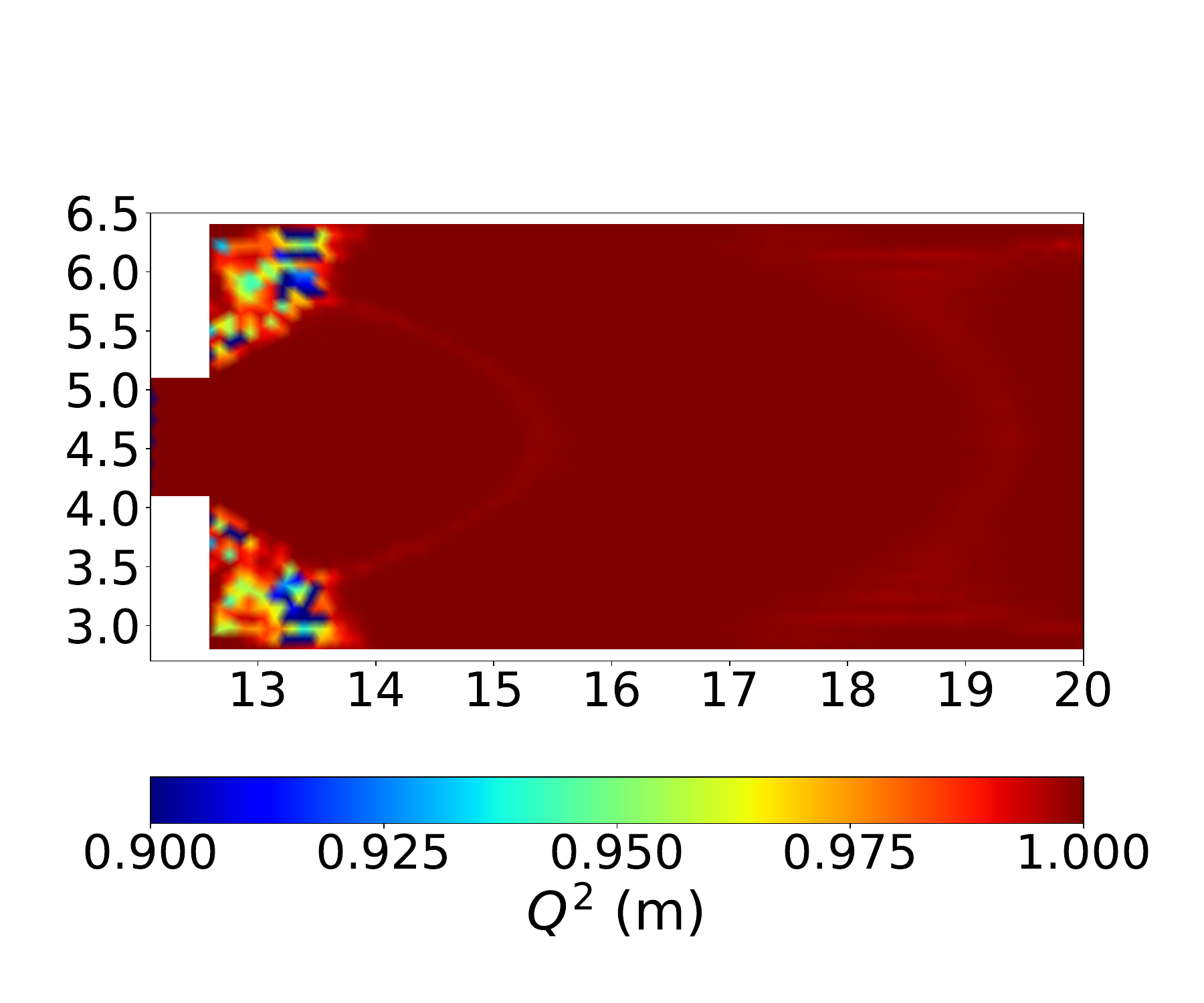}}
\caption{Comparison results between the high fidelity and emulator models on bottom elevation output for the following set of parameters $\zeta_{exp}=\left[\alpha_{MPM}=17.33, \beta_2=0.9, \beta=2.5, \alpha_{ks}=3.8\right]$ respectively (a) and (b) and predictivity criteria $Q^2$ computed on the test subset (c).}
\label{fig:comp_twin_exp}
\end{figure}

Figure \ref{fig:comp_twin_exp} shows the evolution of the bottom elevation prediction, produced by the hydro-morphodynamic TELEMAC-2D/GAIA model, contrasted with its emulator counterpart. The bottom elevation are approximated well by the surrogate model and, as it can be noticed, the absolute difference are negligibles with a MAE (mean absolute error) equals to $1.54\times 10^{-4}$ m and the predictivity criteria is close to the maximum value of $1$ in the major part of the domain. A coefficient close to 1 shows a good fit between the validation database and the result estimated by the reduced order model. Thus, the emulator can be used in place of the high fidelity model in the Markov Chain Monte Carlo algorithm.


\subsection{Bayesian inference}
\label{sub:res:bayes_infer}

As a reminder, the different model parameters $\zeta$ are considered to be independent \textit{a priori} and their uncertainties are assumed to be described by uniform distributions whose limits are defined by the minimum/maximum values of the variation range of each parameter (Section \ref{subsec:parameters}). The prior knowledge on the standard deviation $\sigma_o$ is characterized with a half-normal distribution. In this study, a half-normal prior with a scale parameter of $0.008$, corresponding to the mean value of the standard deviation for the bed elevation measurements \citep{Soares_2012}, is utilized. Using Hamiltonian Monte Carlo sampling, the posterior distributions can be estimated. To ensure robustness, two diagnostics are used in the current study: effective sample size and the split $\hat{R}$ statistic \citep{Gelman_2013}, which check mixing and stationarity of the Markov Chain Monte Carlo (MCMC) sampling. These diagnostics compare variances within and between four Markov chains, each starting from different points in parameter space. After running the MCMC, posterior samples for all parameters are obtained and their marginals are estimated using a standard kernel density estimation (KDE) method. In following, a total of $5\times10^5$ MCMC iterations were found to be sufficient in obtaining the posterior with good resolution.

\subsubsection{Synthetic observations}

Before model inference on complex measurements, twin experiments are conducted to validate the Bayesian framework. These experiments help assess how different assumptions affect the inverse problem methodology. The first twin experiment aims to test the robustness of posterior inference to noise in the observations. An emulator reference simulation with predefined parameters is performed, considering the bottom elevation output as the ``true'' observation field, $y(\bf{x})$. Perturbed versions, $y^o(\bf{x})$, with varying Gaussian noise levels ($\epsilon\sim\mathcal{N}\left(0,\sigma_\epsilon\right)$), emulate imperfect observations. The observations are based on the parameters $\zeta_{exp}=\left[\alpha_{MPM}=17.33, \beta_2=0.9, \beta=2.5, \alpha_{ks}=3.8\right]$. Table \ref{tab:CI:sigma:twin_exp_emulator} presents the expected posterior estimates and their $95$\% credible intervals for different noise levels.


\begin{table}[h]
\caption{95 \% credible interval of morphodynamic parameters for the emulated twin experiment configuration with different noise level}
\label{tab:CI:sigma:twin_exp_emulator}%
\footnotesize
\centering
\begin{tabular}{@{}cccccc@{}}
\toprule
Noise level & \multicolumn{5}{c}{Posterior}\\
$\sigma_\epsilon$& $\alpha_{MPM}$  & $\beta_2$ & $\beta$ & $\alpha_{ks}$ &$\sigma_o$\\
\midrule
$1\times10^{-4}$ & $\left[17.22,17.40\right]$ & $\left[0.896,0.904\right]$  & $\left[2.49,2.51\right]$ & $\left[3.79,3.85\right]$ &  $\left[0.95,1.01\right] \times10^{-4} $\\
$5\times10^{-4}$ & $\left[17.05,17.93\right]$ & $\left[0.88,0.90\right]$  & $\left[2.43,2.53\right]$ & $\left[3.61,3.9\right]$ & $\left[4.85,5.15\right] \times10^{-4} $\\
$1\times10^{-3}$ & $\left[15.77,17.43\right]$ & $\left[0.86,0.91\right]$  & $\left[2.32,2.51\right]$ & $\left[3.74,4.36\right]$ &  $\left[0.98,1.03\right] \times10^{-3} $ \\
$5\times10^{-3}$ & $\left[15.5,24.91\right]$ & $\left[0.78,0.96\right]$  & $\left[1.76,2.76\right]$ & $\left[2.24,4.54\right]$ & $\left[4.8,5.1\right] \times10^{-3} $\\
\bottomrule
\end{tabular}
\end{table}

Table \ref{tab:CI:sigma:twin_exp_emulator} shows that the posterior reliably infers parameter values, with twin experiment values consistently within the $95$\% credible interval. The precision varies with noise levels, demonstrating the efficiency and robustness of posterior inference when observations come from the emulated model. A second twin experiment using synthetic observations from high-fidelity simulation evaluates the emulator's impact on posterior inference robustness. This experiment uses a configuration with Gaussian noise ($\sigma_\epsilon = 1.0 \times 10^{-3}$). The results are similar to those previously presented (Table \ref{tab:CI:sigma:twin_exp_emulator}) with the following $95$ \% credible interval and mean value: $\alpha_{MPM}$, $\left[15.64,17.36\right](16.5)$; $\beta_2$, $\left[0.89,0.94\right](0.91)$; $\beta$, $\left[2.46,2.64\right](2.55)$; $\alpha_{ks}$, $\left[3.77,4.32\right](4.07)$, $\sigma_o$, $\left[0.96,1.0\right] (0.99) \times10^{-3}$. 

To conclude, previously shown comparisons shed light on robustness of the Bayesian procedure based on the neural network emulator to infer morphodynamic parameter values. It can therefore be considered for the investigation on the experimental measurements. 

\subsubsection{Experiment observations}

\begin{figure}[h]
\centering
\includegraphics[width=0.75\textwidth]{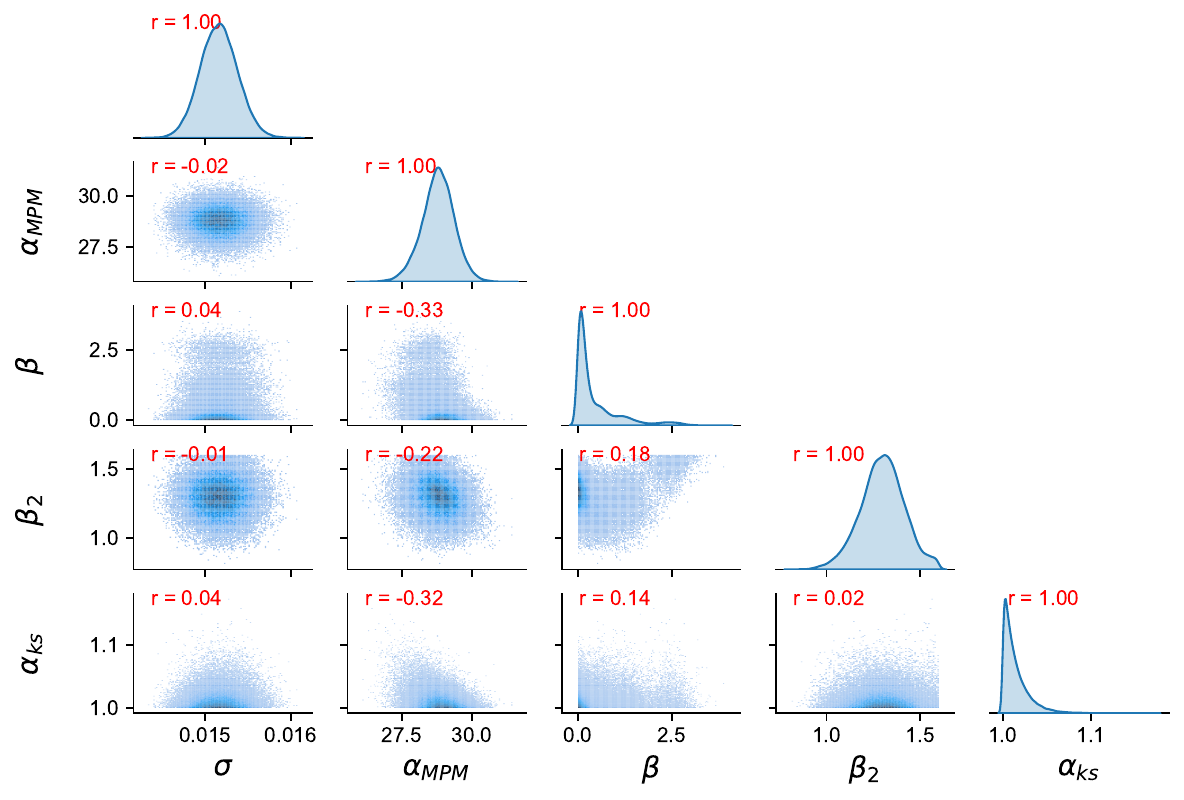}
\caption{Pairwise scatter plots of the posterior sample with $r$ the Pearson correlation coefficient}
\label{fig:pairplot_parameter}
\end{figure}

The observations are the spatial bed elevation measurements taken at the end of the experiment, interpolated onto the mesh computational nodes ranging from $x = 12.78$ to $x = 20.0$. 
As presented in Section \ref{appendix:bayesian}, the inference was performed with an increasing number of observations and replicated several times to evaluate the robustness of the posterior estimates. Figure \ref{fig:pairplot_parameter} displays the posterior samples of the morphodynamic parameters.

Figure \ref{fig:pairplot_parameter} shows that the posterior distributions of the most influential parameters, $\alpha_{MPM}$ and $\alpha_{ks}$, are narrow, with a $95$\% credible interval of about $\pm4$\% of their mean values. In contrast, the $95$\% credible intervals for $\beta$ and $\beta_2$ exceed $\pm15$\%. Additionally, the posterior samples can be considered as not correlated with any other parameters ($|r| < 0.35$). From a physical point of view, the following statements can be observed. The inferred values of the transport coefficient $\alpha_{MPM}$ are consistent with those suggested by \citet{Zech_2008} for modeling the a dam-break front celerity. Regarding the proportional coefficient of the Nikuradse’s equivalent sand-grain roughness $\alpha_{ks}$, the posterior samples align with \citet{van_rijn_1984}. The value of transverse bed slope effect coefficients $\beta_2$ is higher than the original value from \citet{Talmon_1995} and is closer to the calibrated value from \citet{Mendoza_2017}. It appears that with significant morphodynamic changes, $\beta_2$ needs to be set higher than the original value. Lastly, the streamwise bed slope effect coefficient $\beta$ shows the most significant variation in its interval. However, its distribution peaks at very low values, indicating that extreme flow events are driven more by convection than diffusion, as expected.

\begin{figure*}[!h]
\centering
\begin{tabular}{c}
\subfloat[Prior and posterior variability induced by morphodynamic parameters\label{fig:prior_vs_post}]{%
\includegraphics[width=0.75\textwidth]{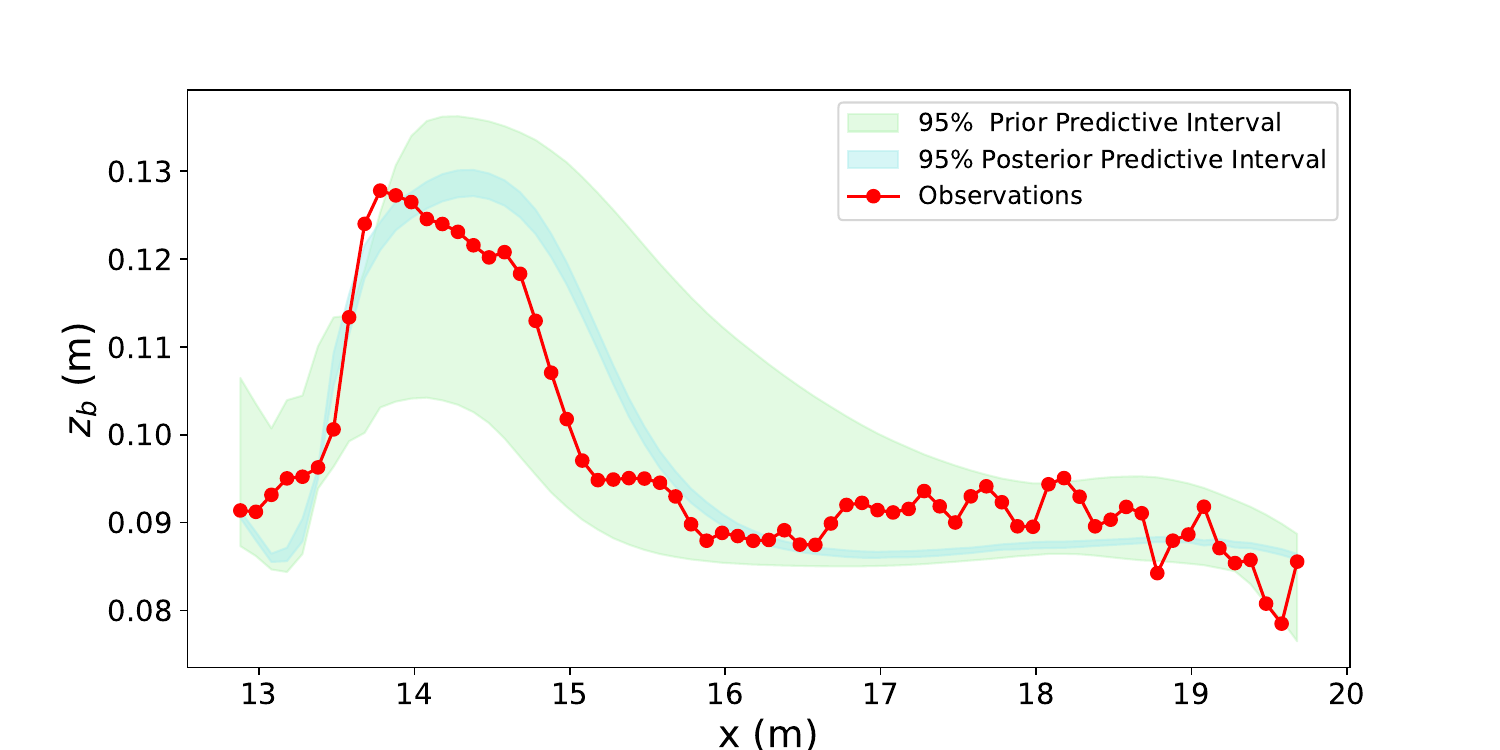}}\\
\subfloat[Total and morphodynamic parameter Posterior variability\label{fig:variability}]{%
\includegraphics[width=0.75\textwidth]{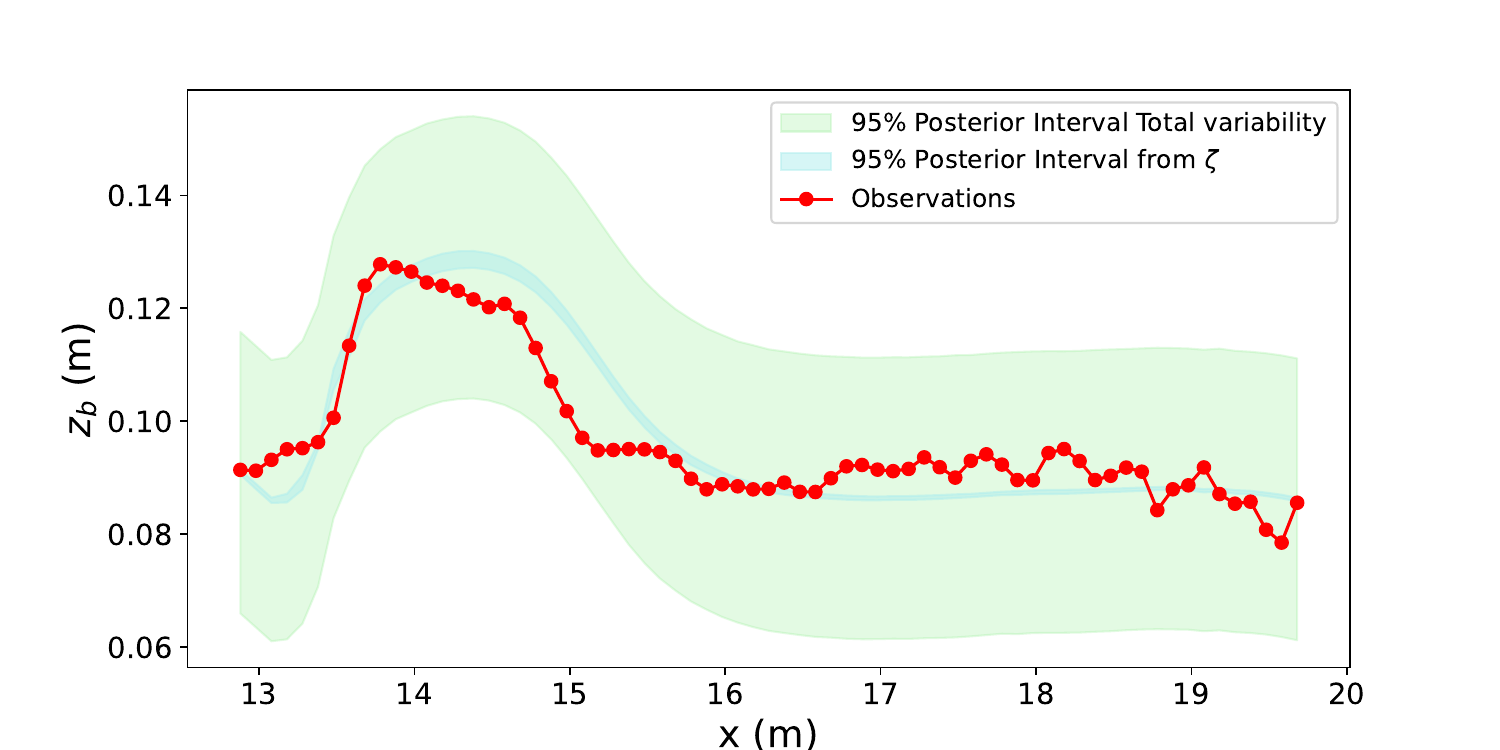}}\\
\end{tabular}
\caption{Posterior Predictive interval on the longitudinal proﬁles for $y = 1.45$ m}
\label{fig:prior_post}
\end{figure*}

Figure \ref{fig:prior_post} displays the comparison between observations and posterior predictive distributions derived from the uncertainty propagation of the inferred morphodynamic parameter. As seen in the Figure, the posterior predictive distribution reproduce accurately the laboratory measurements. In the numerical results, the sediment deposition propagating downstream is consistent with the observed in the laboratory.

Figure \ref{fig:prior_vs_post} shows that introducing data reduces uncertainty in bottom elevation, as seen by the narrower credible posterior interval compared to the prior. Figure \ref{fig:variability} partitions credible posterior predictions to highlight parameter uncertainty and variability from measurement and model errors. The posterior results depend on the calibration scenario. However, the Bayesian framework provide also robust criteria for model selection as demonstrated by \citet{Mohammasi_2018} which can be useful to test different scenarios such as transport and /or observation error models among others.

\section{Discussions}
\label{sec:discussions}

The methodology deployed in this paper aims to incorporate probabilistic methods to enhance the reliability and predictive capabilities of morphodynamic models. 

The application, presented in this article, is a two-dimensional dam-break experiment over a sand bed and is not representative of all the sediment transport study cases. In fact, the dam-break study case is particular in term of alteration intensity in river morphology (high velocities and sands, which are generally easy to mobilize sediments (at least in nature)). This is the main reason why the critical shear stress appears as a non-influential parameter during the sediment transport process. In practice, sediment models are primarily calibrated using the friction coefficient and the sediment median diameter before considering other coefficients. These parameters are not studied in this research as they are known inputs in our case. Ultimately, the value of this work lies less in identifying uncertain parameters and more in developing a robust methodology that can be applied to real-world cases, where the conclusions drawn could be beneficial for further studies. Moreover, the extreme behavior observed in the dam-break scenario over a sand bed should not overshadow the potential discontinuities inherent in sediment transport formulations. The challenging nature of incipient motion necessitates specific methods for uncertainty quantification, as demonstrated in \citet{Bensaid_2024} within the framework of sensitivity analysis.

In addition to the specific study case, the study’s reliance on a specific bedload transport formula introduces limitations related to the physical modeling of sediment transport phenomena. Moreover, the inferred values of sediment bedload magnitude and direction correction factors are not independent of the transport process model used; therefore, the parameter distribution of values found here may result in a very poor fit for a different bedload relation. The use of a generic bedload transport formula covering the existing law variability could be an interesting outlook to overcome this issue. For instance, the law proposed by \citet{Zanke_2020} can be used to encompass around different transport laws from the literature by choosing law parameters with interval of variation covering various transport laws and their validity domains. Since the choice of the transport law could yield varying outcomes, incorporating the transport law parameters into the uncertainty quantification process could reveal that choosing the appropriate law parameters is more critical than optimizing certain process based coefficients.

Having discussed some model limitations, it is now important to address methodological points.

In this work, a Bayesian framework is employed to characterize the inherent parameter uncertainties in a numerical modeling-based morphodynamics application case. To reduce the Bayesian inversion computational cost, a neural network is designed to replicate high-fidelity behavior with a high computational efficiency. To demonstrate the feasibility of the presented methodology, a simple architecture type has been selected in the current work. However, other types of architecture are available \citep{Perez_2024} and the search of the optimal one balancing model complexity with available computational resources must be carried out. The emulator has been created, validated and tested on the basis of learning, validation and test high-fidelity solver computations. Constructing these databases is the most computationally demanding part of the methodology. A major issue concerns the optimization of the number of computational runs needed for sufficient results in the inference process. Approximation with the emulator must become sufficiently accurate with just limited data available. An iterative sampling process to refine locally in the area of interest according to an acquisition function as done in the Bayesian calibration framework could constitute an opportunity to address this issue \citep{Mohammasi_2023}. In practice, High-Performance Computing (HPC) is necessary to tackle realistic applications. If the computational burden is still too high, preliminary experiments may be needed to select the variables for inference, as the amount of obtainable information depends on the problem definition, including the number of variables and the time required for high-fidelity simulations.

\section{Summary and conclusions}
\label{sec:conclusions}

Numerical modeling of morphodynamics is challenging in engineering. Accurate results are crucial for optimizing strategies and reducing costs. However, uncertainties in hydro-morphodynamic modeling arise from inaccurate inputs, model errors, and limited computing resources. Therefore, it's essential to quantify these uncertainties. This paper deployed a step-by-step development of the Bayesian methodology to conduct an uncertainty analysis of $2$D numerical modeling-based morphodynamics representing a dam-break over a sand bed experiment. First, prior knowledge uncertainties are propagated through the dynamical model with the Monte Carlo technique. The Monte Carlo sample allows to estimate the relative influence of each input parameter on the results, helping identify the most relevant parameters and observations for the Bayesian inference process. It also creates a numerical database useful for emulator construction. Finally, with the ingredients defined at the previously stage, a Bayesian framework is employed (i) to characterize the input parameter uncertainty variability taking into account the knowledge a priori, and (ii) to produce probability-based predictions, given observations, in the studied hydro-morphodynamic problem.

Bayesian inference provides a broader framework, enabling the identification of the most probable model from multiple models (derived from different sediment transport laws for instance) given a set of measurements or addressing model uncertainty through model averaging, which weights multiple models based on their posterior probabilities. In addition to the Bayesian framework extension, more diverse and complex hydro-morphodynamic scenarios should be studied to ensure broader applicability and further refinement of predictive models in hydro-morphodynamic studies with the proposed Bayesian framework used here.

\section*{Acknowledgements}

This work was carried out as part of EDF R\&D's research project on river sediment management, whose support the authors gratefully acknowledge. The authors gratefully acknowledge contributions from the open-source community, especially that of the Sensitivity Analysis Library in Python SALib, the open-source machine-learning library PyTorch, the Deep universal probabilistic programming with Python and PyTorch library Pyro and finally, the open source hyperparameter optimization framework to automate hyperparameter search library Optuna.

\section*{Declarations}

\textbf{Competing interests} The authors have no competing interests to declare that are relevant to the content of this article.

\begin{appendices}

\section{Description of the flume sub-zones}
\label{secA1:flume_bathymetry}

As presented in Table \ref{tab:bathy}, the flume experiment used in this study is divided into several sub-zones, each with distinct bathymetric features.

\begin{table}[h]
\caption{Description of the flume sub-zones and their respective bathymetry (including sediments)}
\label{tab:bathy}%
\centering
\begin{tabular}{@{}llll@{}}
\toprule
Area & $x$  & $y$ & bathymetry\\
\midrule
$1$     & $[0.0, 1.76]$     & $[0.0, 9.2]$                   & $-0.10$  \\
$2a$    & $[1.76, 10.59]$   & $[2.8, 3.14]$                  & $-0.155/0.34(y-3.14)$  \\
$2b$    & $[1.76, 10.59]$   & $[3.14, 6.06]$                 & $0.$  \\
$2c$    & $[1.76, 10.59]$   & $[6.06, 6.40]$                 & $-0.155/0.34(y-6.06)$  \\
$3a$    & $[10.59, 10.79]$  & $[2.8, 13.0-0.932x]$           & $-0.155/0.34(y-3.14)$  \\
$3b$    & $[10.59, 10.79]$  & $[13.0-0.932x, 0.932x-3.809]$  & $0.085/0.34(y-3.14)$  \\
$3c$    & $[10.59, 10.79]$  & $[0.932x-3.809, 6.40]$         & $0.155/0.34(y-6.06)$  \\
$4a$    & $[10.79, 11.59]$  & $[2.8, 2.954]$                 & $-0.155/0.34(y-3.14)$  \\
$4b$    & $[10.79, 11.59]$  & $[2.954, 6.246]$               & $0.085$  \\
$4c$    & $[10.79, 11.59]$  & $[6.246, 6.40]$                & $0.155/0.34(y-6.06)$  \\
$5$     & $[11.59, 12.59]$  & $[4.10, 5.10]$                 & $0.085$  \\
$6a$    & $[12.59, 21.09]$  & $[2.8, 2.954]$                 & $-0.155/0.34(y-3.14)$  \\
$6b$    & $[12.59, 21.09]$  & $[2.954, 6.246]$               & $0.085$  \\
$6c$    & $[12.59, 21.09]$  & $[6.246, 6.40]$                & $0.155/0.34(y-6.06)$  \\
$7a$    & $[21.09, 27.59]$  & $[2.8, 3.14]$                  & $-0.155/0.34(y-3.14)$  \\
$7b$    & $[21.09, 27.59]$  & $[3.14, 6.06]$                 & $0.0$  \\
$7c$    & $[21.09, 27.59]$  & $[6.06, 6.40]$                 & $0.155/0.34(y-6.06)$  \\
\bottomrule
\end{tabular}
\end{table}

\section{Summary of the dam-break numerical configuration}
\label{secA2:num_and_phys_config}

Table \ref{tab:num_param} summarizes the main numerical and physical characteristics for the dam-break over a sand bed configuration.

\begin{table}[h]
\caption{Physical and numerical parameters for the dam-break numerical configuration}
\label{tab:num_param}
\centering
\begin{tabular}{@{}ll@{}}
\toprule
\textbf{Hydraulic part} & \\
\hline
Water density & $1000$ kg/m$^{3}$\\
Manning coefficient & $0.0165$ m$^{-1/3}$s  \\
Water viscosity & $10^{-6}$m$^{2}$/s   \\
CFL & less than $0.2$ \\
Simulation time & $20$ s\\
\midrule
\textbf{Sediment part} & \\
\hline
Sediment density & $2630$ kg/m$^{3}$  \\
Sediment diameter & $1.61$ mm \\
Ration coefficient ($\alpha_{ks}$) & $3.$ \\
Sediment Porosity ($\lambda$)& $0.42$    \\
Critical Shield parameter ($\theta_c$)& $0.047$ \\
Bedload transport coefficient ($\alpha_{MPM}$)& $8$ \\
Streamwise bed slope effect ($\beta$)& $1.3$\\
Deviation parameter ($\beta_2$) & $0.85$  \\
\bottomrule
\end{tabular}
\end{table}

\section{Introduction to Hamiltonian mechanics}
\label{sec:uq:subsub:HM}

In Lagrangian mechanics, the equations of motion of a system with $N$ degrees of freedom depend on the coordinates $\left\{q_i\right\}_{i=1,...,N}$ and their corresponding velocity $\left\{\dot{q}_i\right\}_{i=1,...,N}$ and can be written as a function such as $\mathcal{L}\left(q_i,\dot{q}_i,t\right)$. This description of the system state is not the only one. In Hamiltonian mechanics, the momentum variable is introduced such as $p_i = \frac{\partial \mathcal{L}}{\partial \dot{q}_i}$ and the system is described by a function of $\left\{q_i\right\}_{i=1,...,N}$ and $\left\{p_i\right\}_{i=1,...,N}$ known as the Hamiltonian, $\mathcal{H}\left(q_i, p_i,t\right)$. In the following, an isolated system is considered. As a consequence, the $\mathcal{H}$ and $\mathcal{L}$ functions are considered time-independent. From the differential of the Lagrange function $\mathcal{L}$ and the definition of the momentum variable, it can be demonstrated that the quantity $\mathcal{H}\left(q_i,p_i=\frac{\partial \mathcal{L}}{\partial \dot{q}_i},t\right)=\mathcal{E}\left(q_i,\dot{q}_i,t\right)=\sum_{i=1}^{N}p_i\dot{q}_i-\mathcal{L}$ is invariant in time. The quantity $\mathcal{E}\left(q_i,\dot{q}_i,t\right)$ is referred to as the system energy. Finally from the differential of the Hamiltonian expression, the Hamilton’s equations can be deduced.

\begin{equation}
\frac{dq_i}{dt} = \frac{\partial \mathcal{H}}{\partial p_i}, \frac{dp_i}{dt} = -\frac{\partial \mathcal{H}}{\partial q_i}
\label{eq:HM_equations}
\end{equation}

The partial derivatives of the Hamiltonian determine how variables $\textbf{q}$ and $\textbf{p}$ change over time, $t$.  The trajectory of the system in phase space ($\textbf{q}, \textbf{p}$) form a continuous curve such that the Hamiltonian (total energy) remains constant. This implies that the system moves on a surface of constant energy, known as an energy surface or level set of the Hamiltonian.

\section{Sensitivity analysis convergence and robustness}
\label{sec:appendix:as}

When applying sampling-based sensitivity analysis, sensitivity indices are not computed exactly but they are approximated from the available samples. The robustness and convergence of such sensitivity estimates should therefore be assessed \citep{Pianosi_2016}.

\begin{figure*}[!h]
\centering
\begin{tabular}{cc}
\subfloat[P1\label{fig:conv_borgo_us1}]{%
\includegraphics[width=0.5\textwidth]{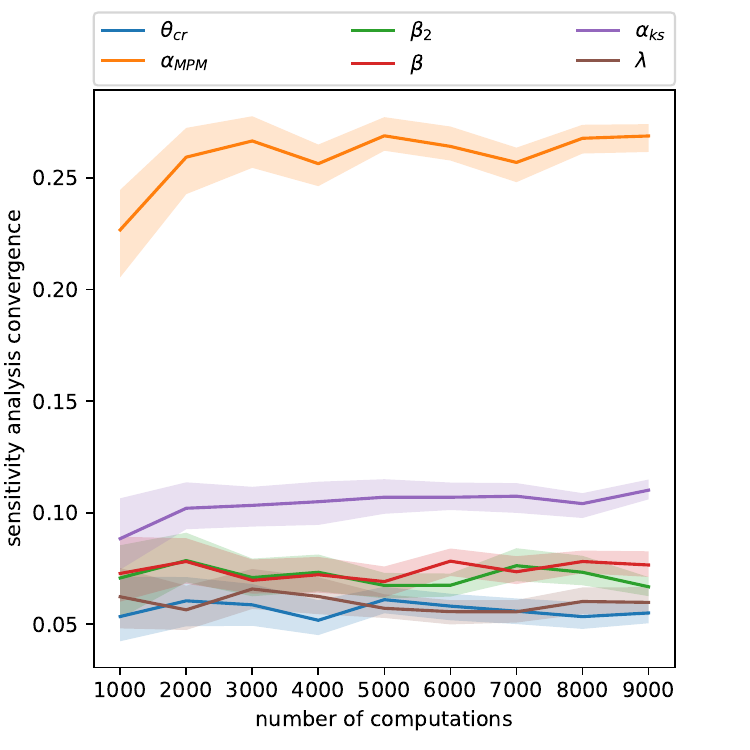}}&
\subfloat[P3\label{fig:conv_borgo_us3}]{%
\includegraphics[width=0.5\textwidth]{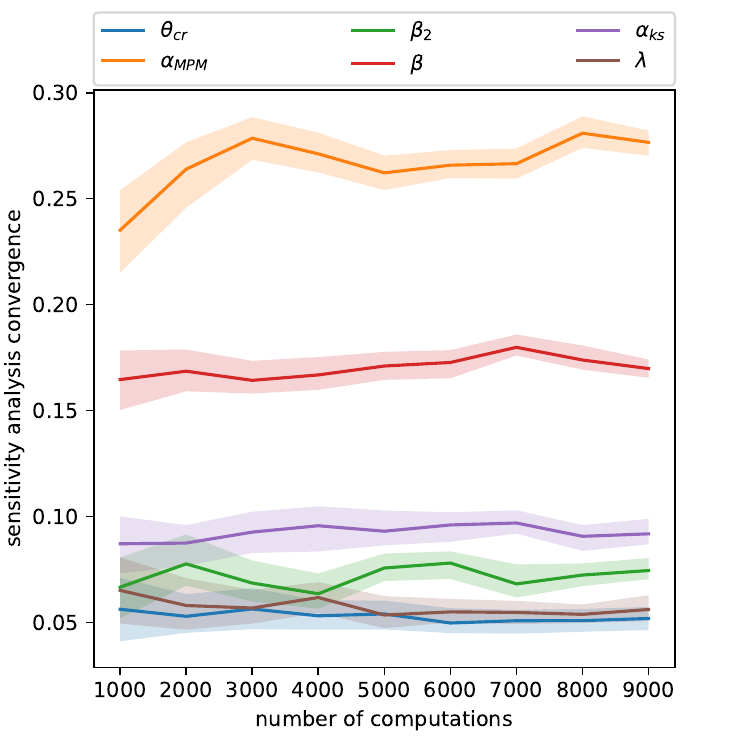}}\\
\subfloat[P6\label{fig:conv_borgo_us6}]{%
\includegraphics[width=0.5\textwidth]{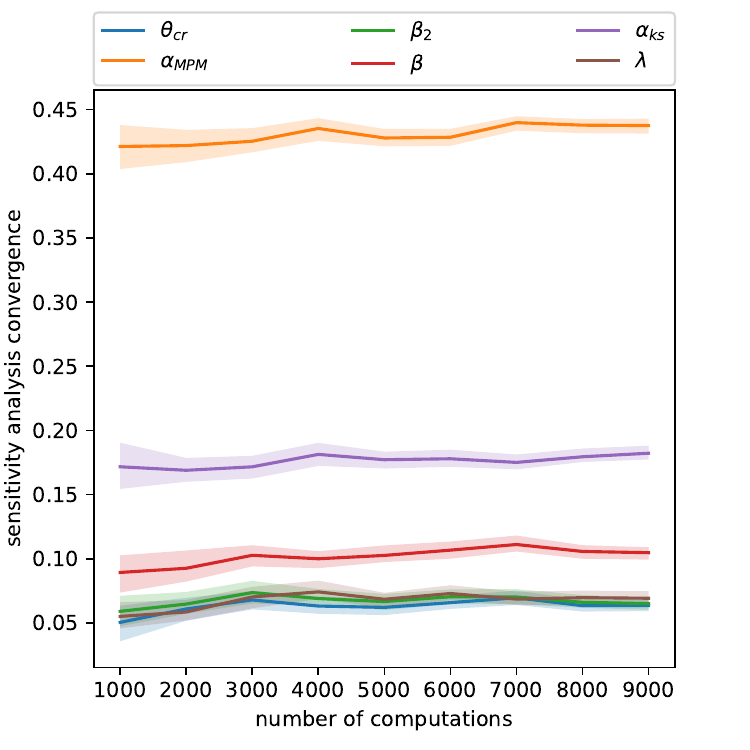}}&
\subfloat[P8\label{fig:conv_borgo_us8}]{%
\includegraphics[width=0.5\textwidth]{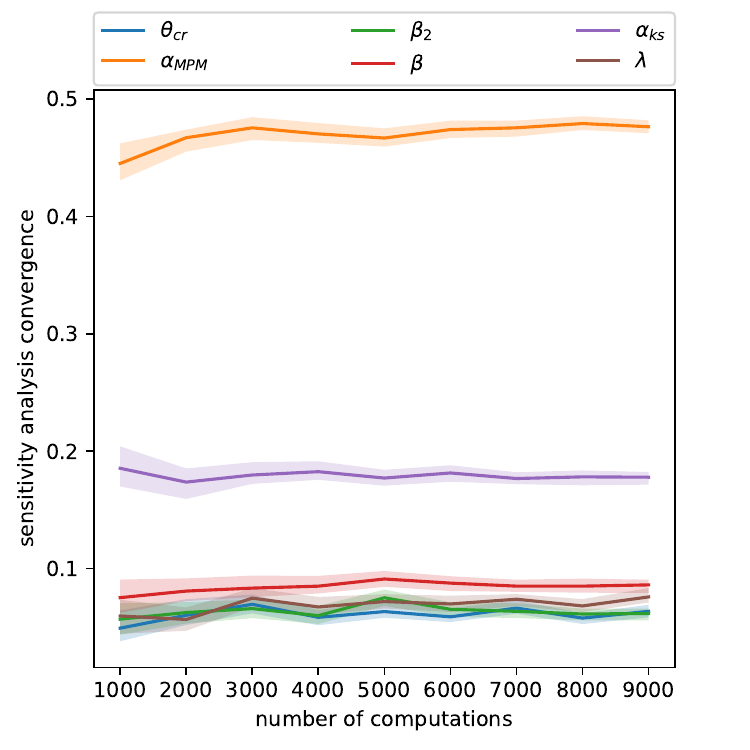}}\\
\end{tabular}
\caption{Convergence graphs of sensitivity indices and their 90\% confidence interval according to sample size, for the final experiment time at control points (a) P1, (b) P2, (c) P3 and (d) P4 (Figure \ref{fig:dam-break_flume})}
\label{fig:convergence_tf_us}
\end{figure*}

To handle these issues, the convergence rates of sensitivity indices are assessed as the sample size increases. A robustness analysis is carried out in order to evaluate the sensitivity of the estimates to the design of experiment. The robustness of the sensitivity indices is analyzed through confidence intervals obtained by repeating 100 times the estimation of Borgonovo sensitivity indices with bootstrap resampling method. Figure \ref{fig:convergence_tf_us} presents the evolution of sensitivity indices obtained at some probe locations at the final experiment time with their respective 90 percent confidence interval.

As displayed on Figure \ref{fig:convergence_tf_us}, the number of samples needed to reach stable sensitivity estimates can vary from one input factor to another and from the location of interest. However, from a sample size of $7000$, the sensitivity indices can be considered as stabilized. As shown in the Figure, the bounds of the $90$ \% confidence intervals are relatively close. Consequently, the estimates can be considered satisfactory in terms of accuracy and robustness. Thus, this number of model evaluation is considered in the investigations.

\newpage

\section{Bayesian inference convergence and robustness}
\label{appendix:bayesian}

A Convergence analysis is carried out to evaluate the impact of observation number on posterior estimates. Table \ref{tab:CI:obs_num:true_exp} summarizes the $95$\% credible interval and mean values of morphodynamic parameters for the flume experiment with varying observation numbers. Observations are randomly taken from the domain of interest.

\begin{table}[h]
\caption{$95$ \% credible interval and mean value of morphodynamic parameters for the flume experiment configuration with different number of considered observations ($N_o$)}
\label{tab:CI:obs_num:true_exp}%
\footnotesize
\centering
\begin{tabular}{@{}cccccc@{}}
\toprule
 & \multicolumn{5}{c}{Posterior}\\
 &  &  & & & \\
 $N_o$& $\alpha_{MPM}$  & $\beta_2$ & $\beta$ & $\alpha_{ks}$ &$\sigma_o\times10^{-2}$\\
\midrule
$500$ & $\left[26.9,31.9\right](30.3)$ & $\left[0.76,1.55\right](1.14)$  & $\left[0.01,1.39\right](0.5)$ & $\left[1.0,1.4\right](1.1)$ &  $\left[1.4,1.6\right] (1.48) $\\
$750$ & $\left[28.4,31.9\right](30.7)$ & $\left[1.11,1.59\right](1.43)$  & $\left[0.006,1.06\right](0.27)$ & $\left[1,1.23\right](1.06)$ & $\left[1.4,1.55\right] (1.48) $\\
$1000$ & $\left[29.8,32.0\right](31.4)$ & $\left[1.22,1.6\right](1.47)$  & $\left[0.006,1.14\right](0.28)$ & $\left[1.,1.27\right](1.09)$ &  $\left[1.4,1.53\right] (1.46) $ \\
$1250$ & $\left[28.6,31.9\right](30.5)$ & $\left[1.08,1.58\right](1.39)$  & $\left[0.005,1.13\right](0.26)$ & $\left[1.,1.12\right](1.03)$ & $\left[1.4,1.52\right] (1.46) $\\
$1500$ & $\left[28.6,31.7\right](30.1)$ & $\left[1.09,1.57\right](1.36)$  & $\left[0.004,0.64\right](0.15)$ & $\left[1.,1.08\right](1.02)$ & $\left[1.4,1.54\right] (1.48) $\\
$1750$ & $\left[28.5,31.4\right](29.9)$ & $\left[1.1,1.55\right](1.35)$  & $\left[0.002,0.43\right](0.11)$ & $\left[1.,1.07\right](1.02)$ & $\left[1.4,1.53\right] (1.47) $\\
$2000$ & $\left[27.6,30.2\right](28.9)$ & $\left[1.08,1.54\right](1.32)$  & $\left[0.005,2.3\right](0.39)$ & $\left[1.,1.064\right](1.02)$ & $\left[1.47,1.57\right] (1.51) $\\
$2250$ & $\left[27.3,29.8\right](28.6)$ & $\left[1.05,1.58\right](1.32)$  & $\left[0.008, 2.86\right](0.87)$ & $\left[1.,1.063\right](1.02)$ & $\left[1.48,1.57\right] (1.53) $\\
$2477$ & $\left[27.6,29.9\right](28.8)$ & $\left[1.06,1.53\right](1.3)$  & $\left[0.006,2.44\right](0.47)$ & $\left[1.,1.054\right](1.01)$ & $\left[1.47,1.56\right] (1.52) $\\
\bottomrule
\end{tabular}
\end{table}

As the number of observations increases, parameter variation decreases. Convergence is reached when credibility intervals and mean values stabilize, which occurs with $2000$ observations. A robustness analysis evaluates the sensitivity of posterior results to the Bayesian algorithm. The $2.5$\% and $97.5$\% quantiles of the parameter posterior distributions are analyzed using confidence intervals, obtained by repeating the methodology $10$ times. Table \ref{tab:true_exp:rob} shows the variation intervals for each inferred parameter.


\begin{table}[h]
\caption{Min-max intervals of the $2.5$\% and $97.5$\% quantiles of the parameter posterior distributions obtained with the maximum number of observations.}
\label{tab:true_exp:rob}%
\footnotesize
\centering
\begin{tabular}{@{}cccccc@{}}
\toprule
Quantile & \multicolumn{5}{c}{Parameters}\\
 & $\alpha_{MPM}$  & $\beta_2$ & $\beta$ & $\alpha_{ks}$ &$\sigma_\epsilon$\\
\midrule
$2.5$ & $\left[27.57,27.58\right]$ & $\left[1.055,1.057\right]$  & $\left[5.8,6.05\right]\times10^{-3}$ & $\left[1.00034,1.00036\right]$ &  $\left[1.474,1.475\right]\times10^{-2}$\\
$97.5$ & $\left[29.89,29.92\right]$ & $\left[1.53,1.54\right]$  & $\left[2.44,2.47\right]$ & $\left[1.053,1.054\right]$ & $\left[1.55,1.56\right]\times10^{-2}$\\
\bottomrule
\end{tabular}
\end{table}

Table \ref{tab:true_exp:rob} shows that the min-max interval bounds are close, indicating the Bayesian procedure with Neural Network produces robust estimates. 

\end{appendices}

\bibliographystyle{abbrvnat}

\begin{thebibliography}{69}
\providecommand{\natexlab}[1]{#1}
\providecommand{\url}[1]{\texttt{#1}}
\expandafter\ifx\csname urlstyle\endcsname\relax
  \providecommand{\doi}[1]{doi: #1}\else
  \providecommand{\doi}{doi: \begingroup \urlstyle{rm}\Url}\fi

\bibitem[Abily et~al.(2016)Abily, Delestre, Bertrand, Duluc, and
  Gourbesville]{Morgan_2016}
M.~Abily, O.~Delestre, N.~Bertrand, C.-M. Duluc, and P.~Gourbesville.
\newblock High-resolution modeling with bi-dimensional shallow water
  equations-based codes – high-resolution topographic data use for flood
  hazard assessment over urban and industrial environments.
\newblock \emph{Procedia Eng.}, 154:\penalty0 853 -- 860, 2016.

\bibitem[Archambeau et~al.(2004)Archambeau, M{\'e}chitoua, and
  Sakiz]{Archambeau_2004}
F.~Archambeau, N.~M{\'e}chitoua, and M.~Sakiz.
\newblock {Code Saturne}: A finite volume code for the computation of turbulent
  incompressible flows - industrial applications.
\newblock \emph{International Journal on Finite Volumes}, 1\penalty0 (1), 2004.

\bibitem[Argaud(2019)]{Argaud_2019}
J.-P. Argaud.
\newblock User documentation, in the salome 9.3 platform, of the {ADAO} module
  for ”data assimilation and optimization”., 2019.

\bibitem[Aronica et~al.(2012)Aronica, Franza, Bates, and Neal]{Aronica_2012}
G.~T. Aronica, F.~Franza, P.~D. Bates, and J.~C. Neal.
\newblock Probabilistic evaluation of flood hazard in urban areas using {Monte
  Carlo} simulation.
\newblock \emph{Hydrol. Process.}, 26\penalty0 (26):\penalty0 3962 -- 3972,
  2012.

\bibitem[Audouin et~al.(2017)Audouin, Goeury, Zaoui, Ata, El~Idrissi~Essebtey,
  Torossian, and Rouge]{Audouin_2017}
Y.~Audouin, C.~Goeury, F.~Zaoui, R.~Ata, S.~El~Idrissi~Essebtey, A.~Torossian,
  and D.~Rouge.
\newblock Interoperability applications of \telemacsystem.
\newblock In \emph{Proc. TELEMAC \& MASCARET User Conf.}, pages 69 -- 76, 2017.

\bibitem[Baudin et~al.(2017)Baudin, Lebrun, Iooss, and Popelin]{Baudin_2017}
M.~Baudin, R.~Lebrun, B.~Iooss, and A.-L. Popelin.
\newblock Openturns: An industrial software for uncertainty quantification in
  simulation.
\newblock \emph{Handbook of Uncertainty Quantification}, pages 2001 -- 2038,
  2017.

\bibitem[Blatman and Sudret(2011)]{BLATMAN20112345}
G.~Blatman and B.~Sudret.
\newblock Adaptive sparse polynomial chaos expansion based on least angle
  regression.
\newblock \emph{J. Comput. Phys.}, 230\penalty0 (6):\penalty0 2345 -- 2367,
  2011.

\bibitem[Braunschweig et~al.(2004)Braunschweig, Leitao, Fernandes, Pina, and
  Nevesn]{Braunschweig_2004}
F.~Braunschweig, P.~C. Leitao, L.~Fernandes, P.~Pina, and R.~J.~J. Nevesn.
\newblock The object-oriented design of the integrated water modeling system
  {MOHID}.
\newblock In \emph{Computational Methods in Water Resources}, volume~55 of
  \emph{Developments in Water Science}, pages 1079 -- 1090. Elsevier, 2004.

\bibitem[Buis et~al.(2006)Buis, Piacentini, and Déclat]{Buis_2006}
S.~Buis, A.~Piacentini, and D.~Déclat.
\newblock Palm: a computational framework for assembling high-performance
  computing applications.
\newblock \emph{Concurr. Comput.}, 18\penalty0 (2):\penalty0 231 -- 245, 2006.

\bibitem[Butler et~al.(2014)Butler, Reed, Fisher-Vanden, Keller, and
  Wagener]{Butler_2014}
M.~P. Butler, P.~M. Reed, K.~Fisher-Vanden, K.~Keller, and T.~Wagener.
\newblock Identifying parametric controls and dependencies in integrated
  assessment models using global sensitivity analysis.
\newblock \emph{Environ. Model. Softw.}, 59:\penalty0 10 -- 29, 2014.

\bibitem[Campbell et~al.(2006)Campbell, McKay, and Williams]{Campbell_2006}
K.~Campbell, M.~D. McKay, and B.~J. Williams.
\newblock Sensitivity analysis when model outputs are functions.
\newblock \emph{Reliab. Eng. Syst. Saf.}, 91\penalty0 (10):\penalty0 1468 --
  1472, 2006.

\bibitem[Carrassi et~al.(2018)Carrassi, Bocquet, Bertino, and
  Evensen]{Carrassi_2018}
A.~Carrassi, M.~Bocquet, L.~Bertino, and G.~Evensen.
\newblock Data assimilation in the geosciences: An overview of methods, issues,
  and perspectives.
\newblock \emph{Wiley Interdiscip. Rev. Clim. Change}, 9\penalty0 (5):\penalty0
  e535, 2018.

\bibitem[Cea et~al.(2011)Cea, Bermúdez, and Puertas]{CEA_2011}
L.~Cea, M.~Bermúdez, and J.~Puertas.
\newblock Uncertainty and sensitivity analysis of a depth-averaged water
  quality model for evaluation of escherichia coli concentration in shallow
  estuaries.
\newblock \emph{Environ. Model. Softw.}, 12:\penalty0 1526 -- 1539, 2011.

\bibitem[Chassignet et~al.(2007)Chassignet, Hurlburt, Smedstad, Halliwell,
  Hogan, Wallcraft, Baraille, and Bleck]{Chassignet_2007}
E.~P. Chassignet, H.~E. Hurlburt, O.~M. Smedstad, G.~R. Halliwell, P.~J. Hogan,
  A.~J. Wallcraft, R.~Baraille, and R.~Bleck.
\newblock The {HYCOM} ({HYbrid Coordinate Ocean Model}) data assimilative
  system.
\newblock \emph{J. Mar. Syst.}, 65\penalty0 (1):\penalty0 60 -- 83, 2007.

\bibitem[Clarke et~al.(1994)Clarke, Glendinning, and Hempel]{Clarke_1994}
L.~Clarke, I.~Glendinning, and R.~Hempel.
\newblock The {MPI Message Passing Interface Standard}.
\newblock In \emph{Proc. Programming Environments for Massively Parallel
  Distributed Systems}, pages 213 -- 218, 1994.

\bibitem[Daloglu et~al.(2014)Daloglu, Nassauer, Riolo, and
  Scavia]{Daloglu_2014}
I.~Daloglu, J.~I. Nassauer, R.~Riolo, and D.~Scavia.
\newblock An integrated social and ecological modeling framework—impacts of
  agricultural conservation practices on water quality.
\newblock \emph{Ecol. Soc.}, 19\penalty0 (3), 2014.

\bibitem[Damblin et~al.(2013)Damblin, Couplet, and Iooss]{Damblin_2013}
G.~Damblin, M.~Couplet, and B.~Iooss.
\newblock Numerical studies of space filling designs: optimization of latin
  hypercube samples and subprojection properties.
\newblock arXiv, 2013.

\bibitem[David et~al.(2013)David, Ascough, Lloyd, Green, Rojas, Leavesley, and
  Ahuja]{David_2013}
O.~David, J.~C. Ascough, W.~Lloyd, T.~R. Green, K.~W. Rojas, G.~H. Leavesley,
  and L.~Ahuja.
\newblock A software engineering perspective on environmental modeling
  framework design: The object modeling system.
\newblock \emph{Environ. Model. Softw.}, 39:\penalty0 201 -- 213, 2013.

\bibitem[DELTARES(2014)]{Deltares_2014}
DELTARES.
\newblock Delft3d-flow user's manual: Simulation of multi-dimensional
  hydrodynamic flows and transport phenomena, including sediments, 2014.

\bibitem[Dung et~al.(2011)Dung, Merz, B\'ardossy, Thang, and Apel]{Dung_2011}
N.~V. Dung, B.~Merz, A.~B\'ardossy, T.~D. Thang, and H.~Apel.
\newblock Multi-objective automatic calibration of hydrodynamic models
  utilizing inundation maps and gauge data.
\newblock \emph{Hydrol. Earth Syst. Sci.}, 15\penalty0 (4):\penalty0 1339 --
  1354, 2011.

\bibitem[Dyhouse et~al.(2007)Dyhouse, Hatchett, and Benn]{dyhouse_2007}
G.~Dyhouse, J.~Hatchett, and J.~Benn.
\newblock \emph{Floodplain Modeling Using HEC-RAS}.
\newblock Haestad Methods water resources modeling collection. Bentley
  Institute Press, 2007.
\newblock ISBN 9781934493021.

\bibitem[Gamboa et~al.(2013)Gamboa, Janon, Klein, and Lagnoux]{Gamboa_2013}
F.~Gamboa, A.~Janon, T.~Klein, and A.~Lagnoux.
\newblock Sensitivity analysis for multidimensional and functional outputs.
\newblock arXiv, 2013.

\bibitem[Gamma et~al.(1994)Gamma, Helm, Johnson, and Vlissides]{Gamma_1994}
E.~Gamma, R.~Helm, R.~Johnson, and J.~M. Vlissides.
\newblock \emph{Design Patterns: Elements of Reusable Object-Oriented
  Software}.
\newblock Addison-Wesley Professional, 1994.

\bibitem[Garcia-Cabrejo and Valocchi(2014)]{Garcia-Cabrejo_20014}
O.~Garcia-Cabrejo and A.~Valocchi.
\newblock Global sensitivity analysis for multivariate output using polynomial
  chaos expansion.
\newblock \emph{Reliab. Eng. Syst. Saf.}, 126:\penalty0 25 -- 36, 2014.

\bibitem[Gil et~al.(2018)Gil, Pierce, Babaie, Banerjee, Borne, Bust, Cheatham,
  Ebert-Uphoff, Gomes, Hill, Horel, Hsu, Kinter, Knoblock, Krum, Kumar,
  Lermusiaux, Liu, North, Pankratius, Peters, Plale, Pope, Ravela, Restrepo,
  Ridley, Samet, Shekhar, Skinner, Smyth, Tikoff, Yarmey, and Zhang]{Gil_2019}
Y.~Gil, S.~A. Pierce, H.~Babaie, A.~Banerjee, K.~Borne, G.~Bust, M.~Cheatham,
  I.~Ebert-Uphoff, C.~Gomes, M.~Hill, J.~Horel, L.~Hsu, J.~Kinter, C.~Knoblock,
  D.~Krum, V.~Kumar, P.~Lermusiaux, Y.~Liu, C.~North, V.~Pankratius, S.~Peters,
  B.~Plale, A.~Pope, S.~Ravela, J.~Restrepo, A.~Ridley, H.~Samet, S.~Shekhar,
  K.~Skinner, P.~Smyth, B.~Tikoff, L.~Yarmey, and J.~Zhang.
\newblock Intelligent systems for geosciences: An essential research agenda.
\newblock \emph{Commun. ACM}, 62\penalty0 (1):\penalty0 76 – 84, 2018.

\bibitem[Goeury et~al.(2015)Goeury, David, Ata, Boyaval, Audouin, Goutal,
  Popelin, Couplet, Baudin, and Barate]{Goeury_2015}
C.~Goeury, T.~David, R.~Ata, S.~Boyaval, Y.~Audouin, N.~Goutal, A.-L. Popelin,
  M.~Couplet, M.~Baudin, and R.~Barate.
\newblock Uncertainty quantification on a real case with \telemacdd.
\newblock In \emph{Proc. TELEMAC \& MASCARET User Conf.}, pages 44 -- 51, 2015.

\bibitem[Goeury et~al.(2017)Goeury, Audouin, and Zaoui]{Goeury_2017}
C.~Goeury, Y.~Audouin, and F.~Zaoui.
\newblock Telapy user's manual, 2017.

\bibitem[Gregersen et~al.(2007)Gregersen, Gijsbers, and Westen]{Gregersen_2007}
J.~B. Gregersen, P.~J.~A. Gijsbers, and S.~J.~P. Westen.
\newblock {OpenMI: Open Modeling Interface}.
\newblock \emph{J. Hydroinformatics}, 9\penalty0 (3):\penalty0 175 -- 191,
  2007.

\bibitem[Griewank and Walther(2008)]{Griewank_2008}
A.~Griewank and A.~Walther.
\newblock \emph{Evaluating Derivatives}.
\newblock Society for Industrial and Applied Mathematics, second edition, 2008.

\bibitem[Harou et~al.(2009)Harou, Pulido-Velazquez, Rosenberg,
  Medellín-Azuara, Lund, and Howitt]{HAROU_2009}
J.~J. Harou, M.~Pulido-Velazquez, D.~E. Rosenberg, J.~Medellín-Azuara, J.~R.
  Lund, and R.~E. Howitt.
\newblock Hydro-economic models: Concepts, design, applications, and future
  prospects.
\newblock \emph{J. Hydrol.}, 375\penalty0 (3):\penalty0 627 -- 643, 2009.

\bibitem[Harpham and Danovaro(2014)]{Harpham_2014}
Q.~Harpham and E.~Danovaro.
\newblock Towards standard metadata to support models and interfaces in a
  hydro-meteorological model chain.
\newblock \emph{J. Hydroinformatics}, 17\penalty0 (2):\penalty0 260 -- 274,
  2014.

\bibitem[Herman et~al.(2015)Herman, Reed, , Zeff, and Characklis]{Herman_2015}
J.~D. Herman, M.~P. Reed, , H.~B. Zeff, and G.~W. Characklis.
\newblock How should robustness be defined for water systems planning under
  change?
\newblock \emph{J. Water Resour. Plan. Manag.}, 141:\penalty0 04015012, 2015.

\bibitem[Hervouet(2007)]{Hervouet_2007}
J.-M. Hervouet.
\newblock \emph{Hydrodynamics of Free Surface Flows : Modeling with the finite
  element method}.
\newblock Wiley, 2007.
\newblock ISBN 978-0470035580.

\bibitem[Huybrechts et~al.(2012)Huybrechts, Villaret, and
  Lyard]{Huybrecht_2012}
N.~Huybrechts, C.~Villaret, and F.~Lyard.
\newblock Optimized predictive two-dimensional hydrodynamic model of the
  {Gironde Estuary} in {France}.
\newblock \emph{J. Waterw. Port. Coast.}, 138\penalty0 (4):\penalty0 312--322,
  2012.

\bibitem[Kim et~al.(2010)Kim, Gerba, and Choi]{Kim_2010}
M.~Kim, C.~P. Gerba, and C.~Y. Choi.
\newblock Assessment of physically-based and data-driven models to predict
  microbial water quality in open channels.
\newblock \emph{J. Environ. Sci}, 22\penalty0 (6):\penalty0 851 -- 857, 2010.

\bibitem[Klein et~al.(2018)Klein, Fort, Sottolichio, Beudin, Mattarolo, Goeury,
  Ponçot, Argaud, Orseau, Tassi, Huybrechts, Cai, Smaoui, Gasset, Nedelec,
  Laborie, Leroux, Barthelemy, Ali, and S.]{Klein_2018}
F.~Klein, A.~Fort, A.~Sottolichio, A.~Beudin, G.~Mattarolo, C.~Goeury,
  A.~Ponçot, J.-P. Argaud, S.~Orseau, P.~Tassi, N.~Huybrechts, S.~Cai,
  H.~Smaoui, R.~Gasset, Y.~Nedelec, V.~Laborie, R.~Leroux, S.~Barthelemy,
  M.~Ali, and K.~S.
\newblock {Gironde XL} : an example of how estuarine scientific research can
  improve dredging and navigation operations in ports.
\newblock In \emph{VII Congreso Nacional de la ATPYC}, 2018.

\bibitem[Knox et~al.(2018)Knox, Meier, Yoon, and Harou]{Knox_2018}
S.~Knox, P.~Meier, J.~Yoon, and J.~J. Harou.
\newblock A {Python} framework for multi-agent simulation of networked resource
  systems.
\newblock \emph{Environ. Model. Softw.}, 103:\penalty0 16 -- 28, 2018.

\bibitem[Knox et~al.(2019)Knox, Tomlinson, Harou, Meier, Rosenberg, Lund, and
  Rheinheimer]{Knox_2019}
S.~Knox, J.~Tomlinson, J.~J. Harou, P.~Meier, D.~E. Rosenberg, J.~R. Lund, and
  D.~E. Rheinheimer.
\newblock An open-source data manager for network models.
\newblock \emph{Environ. Model. Softw.}, 122:\penalty0 104538, 2019.

\bibitem[Lacombe et~al.(2013)Lacombe, Zaoui, and Goutal]{Lacombe_2013}
J.-M. Lacombe, F.~Zaoui, and N.~Goutal.
\newblock Mascaret – an open souce 1-d simulation code for flow hydrodynamic
  with a generic application programming interface.
\newblock \emph{Open Water Journal}, 2\penalty0 (21), 2013.

\bibitem[Lamboni et~al.(2011)Lamboni, Monod, and Makowski]{Lamboni_2014}
M.~Lamboni, H.~Monod, and D.~Makowski.
\newblock Multivariate sensitivity analysis to measure global contribution of
  input factors in dynamic models.
\newblock \emph{Reliab. Eng. Syst. Saf.}, 96\penalty0 (4):\penalty0 450 -- 459,
  2011.

\bibitem[Laniak et~al.(2013)Laniak, Olchin, Goodall, Voinov, Hill, Glynn,
  Whelan, Geller, Quinn, Blind, Peckham, Reaney, Gaber, Kennedy, and
  Hughes]{Laniak_2013}
G.~F. Laniak, G.~Olchin, J.~Goodall, A.~Voinov, M.~Hill, P.~Glynn, G.~Whelan,
  G.~Geller, N.~Quinn, M.~Blind, S.~Peckham, S.~Reaney, N.~Gaber, R.~Kennedy,
  and A.~Hughes.
\newblock Integrated environmental modeling: A vision and roadmap for the
  future.
\newblock \emph{Environ. Model. Softw.}, 39:\penalty0 3 -- 23, 2013.

\bibitem[Lees et~al.(2000)Lees, Camacho, and Chapra]{Lee_2000}
M.~J. Lees, L.~A. Camacho, and S.~Chapra.
\newblock On the relationship of transient storage and aggregated dead zone
  models of longitudinal solute transport in streams.
\newblock \emph{Water Resour. Res.}, 36:\penalty0 213 -- 224, 2000.

\bibitem[Malenovsky et~al.(2012)Malenovsky, Rott, Cihlar, Schaepman,
  García-Santos, Fernandes, and Berger]{Malenovsky_2012}
Z.~Malenovsky, H.~Rott, J.~Cihlar, M.~E. Schaepman, G.~García-Santos,
  R.~Fernandes, and M.~Berger.
\newblock Sentinels for science: Potential of {Sentinel}-1, -2, and -3 missions
  for scientific observations of ocean, cryosphere, and land.
\newblock \emph{Remote Sens. Environ.}, 1120:\penalty0 91 -- 101, 2012.

\bibitem[Marrel et~al.(2017)Marrel, Saint-Geours, and De~Lozzo]{Marrel_2017}
A.~Marrel, N.~Saint-Geours, and M.~De~Lozzo.
\newblock Sensitivity analysis of spatial and/or temporal phenomena.
\newblock \emph{Handbook of Uncertainty Quantification}, pages 1327 -- 1357,
  2017.

\bibitem[Mateus and Neves(2013)]{Mateus_2013}
M.~Mateus and R.~Neves.
\newblock \emph{Ocean modeling for coastal management - Case studies with
  {MOHID}}.
\newblock IST Press, 2013.

\bibitem[Montáns et~al.(2019)Montáns, Chinesta, Gómez-Bombarelli, and
  Kutz]{Montans_2019}
F.~J. Montáns, F.~Chinesta, R.~Gómez-Bombarelli, and J.~N. Kutz.
\newblock Data-driven modeling and learning in science and engineering.
\newblock \emph{Comptes Rendus Mécanique}, 347:\penalty0 845 -- 855, 2019.

\bibitem[Morales and Nocedal(2011)]{Morales_2011}
J.~L. Morales and J.~Nocedal.
\newblock Remark on ``{Algorithm 778: L-BFGS-B: Fortran Subroutines for
  Large-Scale Bound Constrained Optimization}''.
\newblock \emph{ACM Trans. Math. Softw.}, 38\penalty0 (1), 2011.

\bibitem[Morrow et~al.(2019)Morrow, Fu, Ardhuin, Benkiran, Chapron, Cosme,
  d’Ovidio, Farrar, Gille, Lapeyre, Le~Traon, Pascual, Ponte, Qiu, Rascle,
  Ubelmann, Wang, and Zaron]{Morrow_2019}
R.~Morrow, L.-L. Fu, F.~Ardhuin, M.~Benkiran, B.~Chapron, E.~Cosme,
  F.~d’Ovidio, J.~T. Farrar, S.~T. Gille, G.~Lapeyre, P.-Y. Le~Traon,
  A.~Pascual, A.~Ponte, B.~Qiu, N.~Rascle, C.~Ubelmann, J.~Wang, and E.~D.
  Zaron.
\newblock {Global Observations of Fine-Scale Ocean Surface Topography With the
  Surface Water and Ocean Topography (SWOT) Mission}.
\newblock \emph{Front. Mar. Sci.}, 6:\penalty0 232, 2019.

\bibitem[Morvan et~al.(2008)Morvan, Knight, Wright, Tang, and
  Crossley]{Morvan_2008}
H.~Morvan, D.~Knight, N.~Wright, X.~Tang, and A.~Crossley.
\newblock The concept of roughness in fluvial hydraulics and its formulation in
  1d, 2d and 3d numerical simulation models.
\newblock \emph{J. Hydraul. Res.}, 46\penalty0 (2):\penalty0 191 -- 208, 2008.

\bibitem[Nagel et~al.(2020)Nagel, Rieckermann, and Sudret]{Nagel_2020}
J.~B. Nagel, J.~Rieckermann, and B.~Sudret.
\newblock Principal component analysis and sparse polynomial chaos expansions
  for global sensitivity analysis and model calibration: Application to urban
  drainage simulation.
\newblock \emph{Reliab. Eng. Syst. Saf.}, 195:\penalty0 106737, 2020.

\bibitem[Navon(1998)]{Navon_1998}
I.~M. Navon.
\newblock Practical and theoretical aspects of adjoint parameter estimation and
  identifiability in meteorology and oceanography.
\newblock \emph{Dyn. Atmospheres Oceans}, 27\penalty0 (1):\penalty0 55 -- 79,
  1998.

\bibitem[Ogie et~al.(2020)Ogie, Adam, and Perez]{Ogie_2020}
R.~I. Ogie, C.~Adam, and P.~Perez.
\newblock A review of structural approach to flood management in coastal
  megacities of developing nations: current research and future directions.
\newblock \emph{J. Environ. Plan. Manag.}, 63:\penalty0 127 -- 147, 2020.

\bibitem[Orseau et~al.(2020)Orseau, Huybrechts, Tassi, {Pham Van Bang}, and
  Klein]{ORSEAU_2020}
S.~Orseau, N.~Huybrechts, P.~Tassi, D.~{Pham Van Bang}, and F.~Klein.
\newblock Two-dimensional modeling of fine sediment transport with mixed
  sediment and consolidation: Application to the {Gironde Estuary, France}.
\newblock \emph{Int. J. Sediment Res.}, 2020.

\bibitem[Pappenberger et~al.(2008)Pappenberger, Beven, Ratto, and
  Matgen]{Pappenberger_2008}
F.~Pappenberger, K.~J. Beven, M.~Ratto, and P.~Matgen.
\newblock Multi-method global sensitivity analysis of flood inundation models.
\newblock \emph{Adv. Water. Resour.}, 31\penalty0 (1):\penalty0 1 -- 14, 2008.

\bibitem[Peterson(2009)]{Peterson_2009}
P.~Peterson.
\newblock F2py: A tool for connecting {Fortran} and {Python} programs.
\newblock \emph{Int. J. Comput. Sci. Eng.}, 4\penalty0 (4):\penalty0 296 –
  305, 2009.

\bibitem[Pham and Lyard(2012)]{Pham_2012}
C.-T. Pham and F.~Lyard.
\newblock Use of tidal harmonic constants databases to force open boundary
  conditions in {TELEMAC}.
\newblock In \emph{Proc. TELEMAC \& MASCARET User Conf.}, pages 165 -- 172,
  2012.

\bibitem[Pianosi et~al.(2016)Pianosi, Beven, Freer, Hall, Rougier, Stephenson,
  and Wagener]{Pianosi_2016}
F.~Pianosi, K.~Beven, J.~Freer, J.~W. Hall, J.~Rougier, D.~B. Stephenson, and
  T.~Wagener.
\newblock Sensitivity analysis of environmental models: A systematic review
  with practical workflow.
\newblock \emph{Environ. Model. Softw.}, 79:\penalty0 214 -- 232, 2016.

\bibitem[Ran and Nedovic-Budic(2016)]{Ran_2016}
J.~Ran and Z.~Nedovic-Budic.
\newblock Integrating spatial planning and flood risk management: A new
  conceptual framework for the spatially integrated policy infrastructure.
\newblock \emph{Comput. Environ. Urban Syst.}, 57:\penalty0 68 -- 79, 2016.

\bibitem[Ribes and Caremoli(2007)]{Ribes_2007}
A.~Ribes and C.~Caremoli.
\newblock Salome platform component model for numerical simulation.
\newblock In \emph{Proc. International Computer Software and Applications
  Conference}, volume~2, page 553 – 564, 2007.

\bibitem[Saltelli(2002)]{Saltelli_2002}
A.~Saltelli.
\newblock Making best use of model evaluations to compute sensitivity indices.
\newblock \emph{Comput. Phys. Commun.}, 145\penalty0 (2):\penalty0 280 -- 297,
  2002.

\bibitem[Sobol'(2001)]{Sobol_2001}
I.~M. Sobol'.
\newblock Global sensitivity indices for nonlinear mathematical models and
  their {Monte Carlo} estimates.
\newblock \emph{Math. Comput. Simul.}, 55\penalty0 (1):\penalty0 271 -- 280,
  2001.

\bibitem[Solomatine and Ostfeld(2008)]{solomatine_2008}
D.~P. Solomatine and A.~Ostfeld.
\newblock {Data-driven modeling: some past experiences and new approaches}.
\newblock \emph{J. Hydroinformatics}, 10\penalty0 (1):\penalty0 3 -- 22, 2008.

\bibitem[Steiner et~al.(2019)Steiner, Bourinet, and Lahmer]{Steiner_2019}
M.~Steiner, J.-M. Bourinet, and T.~Lahmer.
\newblock An adaptive sampling method for global sensitivity analysis based on
  least-squares support vector regression.
\newblock \emph{Reliab. Eng. Syst. Safe.}, 183:\penalty0 323 -- 340, 2019.

\bibitem[Sudret(2008)]{Sudret2008}
B.~Sudret.
\newblock Global sensitivity analysis using polynomial chaos expansions.
\newblock \emph{Reliab. Eng. Syst. Saf.}, 93\penalty0 (7):\penalty0 964 -- 979,
  2008.

\bibitem[Teng et~al.(2017)Teng, Jakeman, Vaze, Croke, Dutta, and
  Kim]{Teng_2017}
J.~Teng, A.~Jakeman, J.~Vaze, B.~F.~W. Croke, D.~Dutta, and S.~Kim.
\newblock Flood inundation modeling: A review of methods, recent advances and
  uncertainty analysis.
\newblock \emph{Environ. Model. Softw.}, 90:\penalty0 201 -- 216, 2017.

\bibitem[{United States Federal Highway Administration}(1984)]{united1984guide}
{United States Federal Highway Administration}.
\newblock \emph{Guide for Selecting Manning's Roughness Coefficients for
  Natural Channels and Flood Plains}.
\newblock U.S. Department of Transportation, Federal Highway Administration,
  1984.

\bibitem[Van~Rossum and Drake(2009)]{Rossum_2009}
G.~Van~Rossum and F.~L. Drake.
\newblock \emph{Python 3 Reference Manual}.
\newblock CreateSpace, 2009.
\newblock ISBN 1441412697.

\bibitem[Villaret et~al.(2010)Villaret, Van, Huybrechts, Pham~van bang, and
  Boucher]{Villaret_2010}
C.~Villaret, L.~Van, N.~Huybrechts, D.~Pham~van bang, and O.~Boucher.
\newblock Consolidation effects on morphodynamics modeling: application to the
  {Gironde Estuary}.
\newblock \emph{La Houille Blanche}, 6:\penalty0 15 -- 24, 2010.

\bibitem[Zaoui et~al.(2019)Zaoui, Goeury, and Audouin]{Zaoui_2019}
F.~Zaoui, C.~Goeury, and Y.~Audouin.
\newblock A metamodel of the {TELEMAC} errors.
\newblock arXiv, 2019.

\end{thebibliography}


\begin{thebibliography}{75}
\providecommand{\natexlab}[1]{#1}
\providecommand{\url}[1]{\texttt{#1}}
\expandafter\ifx\csname urlstyle\endcsname\relax
  \providecommand{\doi}[1]{doi: #1}\else
  \providecommand{\doi}{doi: \begingroup \urlstyle{rm}\Url}\fi

\bibitem[Akiba et~al.(2019)Akiba, Sano, Yanase, Ohta, and Koyama]{Akiba_2019}
T.~Akiba, S.~Sano, T.~Yanase, T.~Ohta, and M.~Koyama.
\newblock Optuna: A {N}ext-generation {H}yperparameter {O}ptimization {F}ramework, 2019.
\newblock Preprint at \url{https://doi.org/10.48550/arXiv.1907.10902}.

\bibitem[Amoudry and Souza(2011)]{Amoudry_2011}
L.~Amoudry and A.~Souza.
\newblock Deterministic coastal morphological and sediment transport modeling: a review and discussion.
\newblock \emph{Rev. Geophys.}, 49\penalty0 (2), 2011.
\newblock \doi{10.1029/2010RG000341}.

\bibitem[Beckers et~al.(2018)Beckers, Noack, and Wieprecht]{Beckers_2018}
F.~Beckers, M.~Noack, and S.~Wieprecht.
\newblock Uncertainty analysis of a {2D} sediment transport model: an example of the {L}ower {R}iver {S}alzach.
\newblock \emph{J. Soils Sediments.}, 18:\penalty0 3133--–3144, 2018.
\newblock \doi{10.1007/s11368-017-1816-z}.

\bibitem[Beckers et~al.(2020)Beckers, Heredia, Noack, Nowak, Wieprecht, and Oladyshkin]{Beckers_2020}
F.~Beckers, A.~Heredia, M.~Noack, W.~Nowak, S.~Wieprecht, and S.~Oladyshkin.
\newblock Bayesian {C}alibration and {V}alidation of a {L}arge-{S}cale and {T}ime-{D}emanding {S}ediment {T}ransport {M}odel.
\newblock \emph{Water Resour. Res.}, 56\penalty0 (7):\penalty0 e2019WR026966, 2020.
\newblock \doi{10.1029/2019WR026966}.

\bibitem[Ben~Said et~al.(2024)Ben~Said, Alfonsi, Dutfoy, El~Kadi~Abderrezzak, Goeury, Reygner, and Zaoui]{Bensaid_2024}
F.~Ben~Said, A.~Alfonsi, A.~Dutfoy, K.~El~Kadi~Abderrezzak, C.~Goeury, J.~Reygner, and F.~Zaoui.
\newblock Sensitivity analysis in numerical modelling based-morphodynamics, 2024.
\newblock Paper presented at XIXth TELEMAC-MASCARET User Conference, Chambery, FR, 8--10 October 2024.

\bibitem[Betancourt(2018)]{Betancourt_2018}
M.~Betancourt.
\newblock A {C}onceptual {I}ntroduction to {H}amiltonian {M}onte {C}arlo, 2018.
\newblock Preprint at \url{https://arxiv.org/abs/1701.02434}.

\bibitem[Bingham et~al.(2019)Bingham, Chen, Jankowiak, Obermeyer, Pradhan, Karaletsos, Singh, Szerlip, Horsfall, and Goodman]{Bingham_2019}
E.~Bingham, J.~Chen, M.~Jankowiak, F.~Obermeyer, N.~Pradhan, T.~Karaletsos, R.~Singh, P.~Szerlip, P.~Horsfall, and N.~Goodman.
\newblock Pyro: Deep {U}niversal {P}robabilistic {P}rogramming.
\newblock \emph{J. Mach. Learn. Res.}, 20:\penalty0 28:1--28:6, 2019.
\newblock \doi{10.5555/3322706.3322734}.

\bibitem[Borgonovo(2007)]{Borgonovo_2007}
E.~Borgonovo.
\newblock A new uncertainty importance measure.
\newblock \emph{Reliab. Eng. Syst. Saf.}, 92\penalty0 (6):\penalty0 771--784, 2007.
\newblock ISSN 0951-8320.
\newblock \doi{10.1016/j.ress.2006.04.015}.

\bibitem[Buffington and Montgomery(1997)]{Buffington_1997}
J.~Buffington and D.~Montgomery.
\newblock A systematic analysis of eight decades of incipient motion studies, with special reference to gravel-bedded rivers.
\newblock \emph{Water Resour. Res.}, 33\penalty0 (8):\penalty0 1993--2029, 1997.
\newblock \doi{10.1029/96WR03190}.

\bibitem[Cherkassky et~al.(2006)Cherkassky, Krasnopolsky, Solomatine, and Valdes]{Cherkassky_2006}
V.~Cherkassky, V.~Krasnopolsky, D.~Solomatine, and J.~Valdes.
\newblock Computational intelligence in earth sciences and environmental applications: Issues and challenges.
\newblock \emph{Neural Netw.}, 19\penalty0 (2):\penalty0 113--121, 2006.
\newblock \doi{10.1016/j.neunet.2006.01.001}.

\bibitem[Domenico and Schwartz(1998)]{Domenico_1998}
P.~Domenico and F.~Schwartz.
\newblock \emph{Physical and Chemical Hydrogeology}.
\newblock Number 2nd ed. John Wiley \& Sons, Inc., New {Y}ork, 1998.
\newblock ISBN 9780471597629.

\bibitem[Durmus et~al.(2023)Durmus, Gruffaz, Kailas, Saksman, and Vihola]{Durmus_2023}
A.~Durmus, S.~Gruffaz, M.~Kailas, E.~Saksman, and M.~Vihola.
\newblock On the convergence of dynamic implementations of {H}amiltonian {M}onte {C}arlo and {N}o {U}-{T}urn {S}amplers, 2023.
\newblock Preprint at \url{https://arxiv.org/abs/2307.03460}.

\bibitem[Feehan et~al.(2023)Feehan, McCoy, Scheingross, and Gardner]{Feehan_2023}
S.~Feehan, S.~McCoy, J.~Scheingross, and M.~Gardner.
\newblock Quantifying {V}ariability of {I}ncipient-{M}otion {T}hresholds in {G}ravel-{B}edded {R}ivers {U}sing a {G}rain-{S}cale {F}orce-{B}alance {M}odel.
\newblock \emph{J. Geophys. Res. Earth Surf.}, 128\penalty0 (9):\penalty0 e2023JF007162, 2023.
\newblock \doi{10.1029/2023JF007162}.

\bibitem[{Fernandez Luque} and Van~Beek(1976)]{Fernandez_Luque_1976}
R.~{Fernandez Luque} and R.~Van~Beek.
\newblock Erosion {A}nd {T}ransport {O}f {B}ed-{L}oad {S}ediment.
\newblock \emph{J. Hydraul. Res.}, 14\penalty0 (2):\penalty0 127--144, 1976.
\newblock \doi{10.1080/00221687609499677}.

\bibitem[Fortunato et~al.(2009)Fortunato, Bertin, and Oliveira]{Fortunato_2009}
A.~Fortunato, X.~Bertin, and A.~Oliveira.
\newblock Space and time variability of uncertainty in morphodynamic simulations.
\newblock \emph{Coast. Eng.}, 56\penalty0 (8):\penalty0 886 -- 894, 2009.
\newblock \doi{10.1016/j.coastaleng.2009.04.006}.

\bibitem[Garcia(2008)]{Garcia_2008}
M.~Garcia.
\newblock \emph{Sedimentation Engineering}.
\newblock American Society of Civil Engineers, 2008.

\bibitem[Gelman et~al.(2013)Gelman, Carlin, Stern, Dunson, Vehtari, and Rubin]{Gelman_2013}
A.~Gelman, J.~Carlin, H.~Stern, D.~Dunson, A.~Vehtari, and D.~Rubin.
\newblock \emph{Bayesian {D}ata {A}nalysis}.
\newblock Number 3rd ed. Chapman and Hall/CRC, New York, 2013.

\bibitem[Goeury et~al.(2022)Goeury, Audouin, and Zaoui]{Goeury_2022}
C.~Goeury, Y.~Audouin, and F.~Zaoui.
\newblock Interoperability and computational framework for simulating open channel hydraulics: Application to sensitivity analysis and calibration of gironde estuary model.
\newblock \emph{Environ. Model. Softw.}, 148:\penalty0 105243, 2022.
\newblock ISSN 1364-8152.
\newblock \doi{10.1016/j.envsoft.2021.105243}.

\bibitem[Goldstein et~al.(2019)Goldstein, Coco, and Plant]{Goldstein_2019}
E.~B. Goldstein, G.~Coco, and G.~P. Plant.
\newblock A review of machine learning applications to coastal sediment transport and morphodynamics.
\newblock \emph{Earth-Sci. Rev.}, 194:\penalty0 97--108, 2019.
\newblock ISSN 0012-8252.
\newblock \doi{10.1016/j.earscirev.2019.04.022}.

\bibitem[Gomez and Phillips(1999)]{Gomez_1999}
B.~Gomez and J.~Phillips.
\newblock {Deterministic {U}ncertainty in {B}ed {L}oad {T}ransport}.
\newblock \emph{J. Hydraul. Eng.}, 20\penalty0 (3), 03 1999.
\newblock \doi{10.1061/(ASCE)0733-9429(1999)125:3(305)}.

\bibitem[Goulart et~al.(2023)Goulart, Bleninger, {de Oliveira Fagundes}, and Mainardi]{Goulart_2023}
C.~Goulart, T.~Bleninger, U.~{de Oliveira Fagundes}, and F.~Mainardi.
\newblock Modeling uncertainties of reservoir flushing simulations.
\newblock \emph{Int. J. Sediment Res.}, 38\penalty0 (5):\penalty0 698--710, 2023.
\newblock ISSN 1001-6279.
\newblock \doi{10.1016/j.ijsrc.2023.04.005}.

\bibitem[Hastings(1970)]{Hastings_1970}
W.~Hastings.
\newblock Monte {C}arlo {S}ampling {M}ethods {U}sing {M}arkov {C}hains and {T}heir {A}pplications.
\newblock \emph{Biometrika}, 57\penalty0 (1):\penalty0 97--109, 1970.
\newblock \doi{10.2307/2334940}.

\bibitem[Herman and Usher(2017)]{Herman_2017}
J.~Herman and W.~Usher.
\newblock Salib: An open-source {P}ython library for {S}ensitivity {A}nalysis.
\newblock \emph{J. Open Source Softw.}, 2, 01 2017.
\newblock \doi{10.21105/joss.00097}.

\bibitem[Hervouet(2007)]{Hervouet_2007}
J.-M. Hervouet.
\newblock \emph{{H}ydrodynamics of {F}ree {S}urface {F}lows: {M}odelling With the {F}inite {E}lement Method}.
\newblock John Wiley \& Sons, Ltd, 2007.

\bibitem[Hoffman and Gelman(2014)]{Hoffman_2014}
M.~D. Hoffman and A.~Gelman.
\newblock The {N}o-{U}-turn sampler: adaptively setting path lengths in {H}amiltonian {M}onte {C}arlo.
\newblock \emph{J. Mach. Learn. Res.}, 15\penalty0 (1):\penalty0 1593–--1623, jan 2014.
\newblock ISSN 1532-4435.

\bibitem[Hou et~al.(2018)Hou, Han, Li, Guo, and Qin]{Hou_2018}
J.~Hou, H.~Han, Z.~Li, K.~Guo, and Y.~Qin.
\newblock {Effects of morphological change on fluvial flood patterns evaluated by a hydro-geomorphological model}.
\newblock \emph{J. Hydroinformatics.}, 20\penalty0 (3):\penalty0 633--644, 02 2018.
\newblock \doi{10.2166/hydro.2018.142}.

\bibitem[Hu and Guo(2011)]{Hu_2011}
C.~Hu and Q.~Guo.
\newblock Near-{B}ed {S}ediment {C}oncentration {D}istribution and {B}asic {P}robability of {S}ediment {M}ovement.
\newblock \emph{J. Hydraul. Eng.}, 137\penalty0 (10):\penalty0 1269--1275, 10 2011.
\newblock \doi{10.1061/(ASCE)HY.1943-7900.0000382}.

\bibitem[Kazhyken et~al.(2024)Kazhyken, Valseth, Videman, and Dawson]{Kazhyken_2024}
K.~Kazhyken, E.~Valseth, J.~Videman, and C.~Dawson.
\newblock Application of a dispersive wave hydro-sediment-morphodynamic model in the ria formosa lagoon.
\newblock \emph{Comput. Geosci.}, 28\penalty0 (10):\penalty0 1031--–1047, 2024.
\newblock \doi{0.1007/s10596-024-10305-x}.

\bibitem[Koch and Flokstra(1980)]{Koch_1980}
F.~Koch and C.~Flokstra.
\newblock Bed level computations for curved alluvial channels, 1980.
\newblock Paper presented at the 19th IAHR Congress, New Delhi, India, February, 1981.

\bibitem[Kopmann et~al.(2012)Kopmann, Merkel, and Riehme]{kopmann_2012}
R.~Kopmann, U.~Merkel, and J.~Riehme.
\newblock Using reliability analysis in morphodynamic simulation with {TELEMAC-2D/SISYPHE}, 2012.
\newblock Paper presented at XIXth TELEMAC-MASCARET User Conference, Oxford, UK, 18--19 October 2012.

\bibitem[Leisenring and Moradkhani(2012)]{Leisenring_2012}
M.~Leisenring and H.~Moradkhani.
\newblock Analyzing the uncertainty of suspended sediment load prediction using sequential data assimilation.
\newblock \emph{J. Hydrol.}, 468-469:\penalty0 268--282, 2012.
\newblock ISSN 0022-1694.
\newblock \doi{10.1016/j.jhydrol.2012.08.049}.

\bibitem[Lu et~al.(2017)Lu, Pu, Wang, Hu, and Wang]{Lu_2017}
Z.~Lu, H.~Pu, F.~Wang, Z.~Hu, and L.~Wang.
\newblock The expressive power of neural networks: A view from the width, 2017.
\newblock Preprint at \url{https://doi.org/10.48550/arXiv.1709.02540}.

\bibitem[Mendoza et~al.(2017)Mendoza, Abad, Langendoen, Wang, Tassi, and El~Kadi~Abderrezzak]{Mendoza_2017}
A.~Mendoza, J.~Abad, E.~Langendoen, D.~Wang, P.~Tassi, and K.~El~Kadi~Abderrezzak.
\newblock Effect of {S}ediment {T}ransport {B}oundary {C}onditions on the {N}umerical {M}odeling of {B}ed {M}orphodynamics.
\newblock \emph{J. Hydraul. Eng.}, 143\penalty0 (4):\penalty0 04016099, 2017.
\newblock \doi{10.1061/(ASCE)HY.1943-7900.0001208}.

\bibitem[Meyer-Peter and M\"uller(1948)]{Meyer_1948}
E.~Meyer-Peter and R.~M\"uller.
\newblock Formulas for {B}ed-{L}oad transport, 1948.
\newblock Paper presented at the 2nd Meeting, IAHR, Stockholm, Sweden, 1948.

\bibitem[Mezbache et~al.(2020)Mezbache, Paquier, and Hasbaia]{Mezbache_2020}
S.~Mezbache, A.~Paquier, and M.~Hasbaia.
\newblock A coupled {1-D/2-D} model for simulating river sediment transport and bed evolution.
\newblock \emph{J. Hydroinformatics.}, 125\penalty0 (3):\penalty0 1122--1137, 06 2020.
\newblock \doi{10.2166/hydro.2020.020}.

\bibitem[Mohammadi et~al.(2018)Mohammadi, Kopmann, Guthke, Oladyshkin, and Nowak]{Mohammasi_2018}
F.~Mohammadi, R.~Kopmann, A.~Guthke, S.~Oladyshkin, and W.~Nowak.
\newblock Bayesian selection of hydro-morphodynamic models under computational time constraints.
\newblock \emph{Adv. Water Resour.}, 117:\penalty0 53--64, 2018.
\newblock ISSN 0309-1708.
\newblock \doi{10.1016/j.advwatres.2018.05.007}.

\bibitem[Mohammadi et~al.(2023)Mohammadi, Eggenweiler, Flemisch, Oladyshkin, Rybak, Schneider, and Weishaupt]{Mohammasi_2023}
F.~Mohammadi, E.~Eggenweiler, B.~Flemisch, S.~Oladyshkin, I.~Rybak, M.~Schneider, and K.~Weishaupt.
\newblock A surrogate-assisted uncertainty-aware bayesian validation framework and its application to coupling free flow and porous-medium flow.
\newblock \emph{Comput. Geosci.}, 27:\penalty0 663--686, 2023.
\newblock \doi{10.1007/s10596-023-10228-z}.

\bibitem[Morvan et~al.(2008)Morvan, Knight, Wright, Tang, and Crossley]{Morvan_2008}
H.~Morvan, D.~Knight, N.~Wright, X.~Tang, and A.~Crossley.
\newblock The concept of roughness in fluvial hydraulics and its formulation in {1D}, {2D} and {3D} numerical simulation models.
\newblock \emph{J. Hydraul. Res.}, 46:\penalty0 191--208, 03 2008.
\newblock \doi{10.1080/00221686.2008.9521855}.

\bibitem[Mouradi et~al.(2016)Mouradi, Audouin, Goeury, Claude, Tassi, and El~Kadi~Abderrezzak]{Mouradi_2016}
R.-S. Mouradi, Y.~Audouin, C.~Goeury, N.~Claude, P.~Tassi, and K.~El~Kadi~Abderrezzak.
\newblock Sensitivity analysis and uncertainty quantification in 2d morphodynamic models using a newly implemented {API} for {TELEMAC2D/SISYPHE}, 2016.
\newblock Paper presented at XXIIIth TELEMAC-MASCARET User Conference, Paris, FR, 11--13 October 2016.

\bibitem[Mouris et~al.(2023)Mouris, Acuna, Schwindt, Mohammadi, Haun, Wieprecht, and Oladyshkin]{Mouris_2023}
K.~Mouris, E.~Acuna, S.~Schwindt, F.~Mohammadi, S.~Haun, S.~Wieprecht, and S.~Oladyshkin.
\newblock Stability criteria for {B}ayesian calibration of reservoir sedimentation models.
\newblock \emph{Model. Earth Syst. Environ.}, 9:\penalty0 1--19, 02 2023.
\newblock \doi{10.1007/s40808-023-01712-7}.

\bibitem[M\"uller and Hassan(2018)]{Muller_2018}
T.~M\"uller and M.~A. Hassan.
\newblock Fluvial response to changes in the magnitude and frequency of sediment supply in a {1-D} model.
\newblock \emph{Earth Surf. Dyn.}, 6\penalty0 (4):\penalty0 1041--1057, 2018.
\newblock \doi{10.5194/esurf-6-1041-2018}.

\bibitem[Neal(2011)]{Neal_2012}
R.~Neal.
\newblock {MCMC} using {H}amiltonian dynamics.
\newblock In S.~Brooks, A.~Gelman, G.~Jones, and X.-L. Meng, editors, \emph{Handbook of Markov Chain Monte Carlo}, pages 113--162. Chapman and Hall/CRC, New {Y}ork, 2011.

\bibitem[Oliveira et~al.(2021)Oliveira, Ballio, and Maia]{Oliveira_2021}
B.~Oliveira, F.~Ballio, and R.~Maia.
\newblock Numerical modelling-based sensitivity analysis of fluvial morphodynamics.
\newblock \emph{Environ. Model. Softw.}, 135:\penalty0 104903, 2021.
\newblock \doi{10.1016/j.envsoft.2020.104903}.

\bibitem[Orseau et~al.(2021)Orseau, Huybrechts, Tassi, {Pham Van Bang}, and Klein]{Orseau_2021}
S.~Orseau, N.~Huybrechts, P.~Tassi, D.~{Pham Van Bang}, and F.~Klein.
\newblock Two-dimensional modeling of fine sediment transport with mixed sediment and consolidation: Application to the gironde estuary, france.
\newblock \emph{Int. J. Sediment Res.}, 36\penalty0 (6):\penalty0 736--746, 2021.
\newblock ISSN 1001--6279.
\newblock \doi{10.1016/j.ijsrc.2019.12.005}.

\bibitem[Parker et~al.(2003)Parker, Seminara, and Solari]{Parker_2003}
G.~Parker, G.~Seminara, and L.~Solari.
\newblock Bed load at low shields stress on arbitrarily sloping beds: Alternative entrainment formulation.
\newblock \emph{Water Resour. Res.}, 39\penalty0 (7), 2003.
\newblock \doi{10.1029/2001WR001253}.

\bibitem[Paszke et~al.(2019)Paszke, Gross, Massa, Lerer, Bradbury, Chanan, Killeen, Lin, Gimelshein, Antiga, Desmaison, Köpf, Yang, DeVito, Raison, Tejani, Chilamkurthy, Steiner, Fang, Bai, and Chintala]{Paszke_2019}
A.~Paszke, S.~Gross, F.~Massa, A.~Lerer, J.~Bradbury, G.~Chanan, T.~Killeen, Z.~Lin, N.~Gimelshein, L.~Antiga, A.~Desmaison, A.~Köpf, E.~Yang, Z.~DeVito, M.~Raison, A.~Tejani, S.~Chilamkurthy, B.~Steiner, L.~Fang, J.~Bai, and S.~Chintala.
\newblock Pytorch: An {I}mperative {S}tyle, {H}igh-{P}erformance {D}eep {L}earning {L}ibrary, 2019.
\newblock Preprint at \url{https://arxiv.org/abs/1912.01703}.

\bibitem[Perez and Poncet(2024)]{Perez_2024}
S.~Perez and P.~Poncet.
\newblock Auto-weighted bayesian physics-informed neural networks and robust estimations for multitask inverse problems in pore-scale imaging of dissolution.
\newblock \emph{Comput. Geosci.}, 28:\penalty0 1175--–1215, 2024.
\newblock \doi{10.1007/s10596-024-10313-x}.

\bibitem[Pianforini et~al.(2024)Pianforini, Dazzi, Pilzer, and Vacondio]{Pianforini_2024}
M.~Pianforini, S.~Dazzi, A.~Pilzer, and R.~Vacondio.
\newblock Real-time flood maps forecasting for dam-break scenarios with a transformer-based deep learning model.
\newblock \emph{J. Hydrol.}, 635:\penalty0 131169, 2024.
\newblock ISSN 0022-1694.
\newblock \doi{10.1016/j.jhydrol.2024.131169}.

\bibitem[Pianosi et~al.(2016)Pianosi, Beven, Freer, Hall, Rougier, Stephenson, and Wagener]{Pianosi_2016}
F.~Pianosi, K.~Beven, J.~Freer, J.~Hall, J.~Rougier, D.~Stephenson, and T.~Wagener.
\newblock Sensitivity analysis of environmental models: A systematic review with practical workflow.
\newblock \emph{Environ. Model. Softw.}, 79:\penalty0 214 -- 232, 2016.
\newblock ISSN 1364-8152.
\newblock \doi{10.1016/j.envsoft.2016.02.008}.

\bibitem[Pinto et~al.(2006)Pinto, Fortunato, and Freire]{Pinto_2006}
L.~Pinto, A.~Fortunato, and P.~Freire.
\newblock Sensitivity analysis of non-cohesive sediment transport formulae.
\newblock \emph{Cont. Shelf Res.}, 26\penalty0 (15):\penalty0 1826--1839, 2006.
\newblock \doi{10.1016/j.csr.2006.06.001}.

\bibitem[Pittaluga et~al.(2015)Pittaluga, Tambroni, Canestrelli, Slingerland, Lanzoni, and Seminara]{Pittaluga_2015}
M.~Pittaluga, N.~Tambroni, A.~Canestrelli, R.~Slingerland, S.~Lanzoni, and G.~Seminara.
\newblock Where river and tide meet: The morphodynamic equilibrium of alluvial estuaries.
\newblock \emph{J. Geophys. Res. Earth Surf.}, 120\penalty0 (1):\penalty0 75--94, 2015.
\newblock \doi{10.1002/2014JF003233}.

\bibitem[Plischke et~al.(2013)Plischke, Borgonovo, and Smith]{Plischke_2013}
E.~Plischke, E.~Borgonovo, and C.~Smith.
\newblock Global sensitivity measures from given data.
\newblock \emph{Eur. J. Oper. Res.}, 226\penalty0 (3):\penalty0 536--550, 2013.
\newblock ISSN 0377-2217.
\newblock \doi{10.1016/j.ejor.2012.11.047}.

\bibitem[Saltelli(2002)]{Saltelli_2002}
A.~Saltelli.
\newblock Making best use of model evaluations to compute sensitivity indices.
\newblock \emph{Comput. Phys. Commun.}, 145\penalty0 (2):\penalty0 280 -- 297, 2002.
\newblock ISSN 0010-4655.
\newblock \doi{10.1016/S0010-4655(02)00280-1}.

\bibitem[Samantaray and Ghose(2018)]{Samantaray_2018}
S.~Samantaray and D.~Ghose.
\newblock Evaluation of suspended sediment concentration using descent neural networks.
\newblock \emph{Procedia Comput. Sci.}, 132:\penalty0 1824--1831, 2018.
\newblock ISSN 1877-0509.
\newblock \doi{10.1016/j.procs.2018.05.138}.

\bibitem[Schmelter et~al.(2012)Schmelter, Erwin, and Wilcock]{Schmelter_2012}
M.~Schmelter, S.~Erwin, and P.~Wilcock.
\newblock Accounting for uncertainty in cumulative sediment transport using bayesian statistics.
\newblock \emph{Geomorphology}, 175-176:\penalty0 1--13, 2012.
\newblock ISSN 0169-555X.
\newblock \doi{10.1016/j.geomorph.2012.06.012}.

\bibitem[Sengupta et~al.(2020)Sengupta, Basak, Saikia, Paul, Tsalavoutis, Atiah, Ravi, and Peters]{Sengupta_2020}
S.~Sengupta, S.~Basak, P.~Saikia, S.~Paul, V.~Tsalavoutis, F.~Atiah, V.~Ravi, and A.~Peters.
\newblock A review of deep learning with special emphasis on architectures, applications and recent trends.
\newblock \emph{Knowl.-Based Syst.}, 194:\penalty0 105596, 2020.
\newblock ISSN 0950-7051.
\newblock \doi{10.1016/j.knosys.2020.105596}.

\bibitem[Siedersleben et~al.(2021)Siedersleben, Jocham, Aufleger, and Klar]{Siedersleben_2021}
J.~Siedersleben, S.~Jocham, M.~Aufleger, and R.~Klar.
\newblock Morphodynamic {M}odelling with {U}ncertain {G}eometry {I}nput.
\newblock \emph{Water}, 13\penalty0 (16), 2021.
\newblock \doi{10.3390/w13162248}.

\bibitem[Soares-Fraz\~{a}o et~al.(2012)Soares-Fraz\~{a}o, Canelas, Cao, Cea, Chaudhry, Die~Moran, El~Kadi~Abderrezzak, Ferreira, Cadórniga, Gonzalez-Ramirez, Greco, Huang, Imran, Le~Coz, Marsooli, Paquier, Pender, Pontillo, Puertas, Spinewine, Swartenbroekx, Tsubaki, Villaret, Wu, Yue, and Zech]{Soares_2012}
S.~Soares-Fraz\~{a}o, R.~Canelas, Z.~Cao, L.~Cea, H.~Chaudhry, A.~Die~Moran, K.~El~Kadi~Abderrezzak, R.~Ferreira, I.~Cadórniga, N.~Gonzalez-Ramirez, M.~Greco, W.~Huang, J.~Imran, J.~Le~Coz, R.~Marsooli, A.~Paquier, G.~Pender, M.~Pontillo, J.~Puertas, B.~Spinewine, C.~Swartenbroekx, R.~Tsubaki, C.~Villaret, W.~Wu, Z.~Yue, and Y.~Zech.
\newblock Dam-break flows over mobile beds: experiments and benchmark tests for numerical models.
\newblock \emph{J. Hydraul. Res.}, 50\penalty0 (4):\penalty0 364--375, 2012.
\newblock \doi{10.1080/00221686.2012.689682}.

\bibitem[Struiksma and Crosato(1989)]{Struiksma_1989}
N.~Struiksma and A.~Crosato.
\newblock Analysis of a {2-D} {B}ed {T}opography {M}odel for {R}ivers.
\newblock In S.~Ikeda and G.~Parker, editors, \emph{River Meandering}, pages 153--180. American Geophysical Union (AGU), New {Y}ork, 1989.

\bibitem[Struiksma et~al.(1985)Struiksma, Olesen, Flokstra, and De~Vriend]{Struiksma_1985}
N.~Struiksma, K.~W. Olesen, C.~Flokstra, and H.~J. De~Vriend.
\newblock Bed deformation in curved alluvial channels.
\newblock \emph{J. Hydraul. Res.}, 23\penalty0 (1):\penalty0 57--79, 1985.
\newblock \doi{10.1080/00221688509499377}.

\bibitem[Talmon et~al.(1995)Talmon, Struiksma, and Van~Mierlo]{Talmon_1995}
A.~Talmon, N.~Struiksma, and M.~Van~Mierlo.
\newblock Laboratory measurements of the direction of sediment transport on transverse alluvial-bed slopes.
\newblock \emph{J. Hydraul. Res.}, 33\penalty0 (4):\penalty0 495--517, 1995.
\newblock \doi{10.1080/00221689509498657}.

\bibitem[Tassi et~al.(2023)Tassi, Benson, Delinares, Fontaine, Huybrechts, Kopmann, Pavan, Pham, Taccone, and Walther]{Tassi_2023}
P.~Tassi, T.~Benson, M.~Delinares, J.~Fontaine, N.~Huybrechts, R.~Kopmann, S.~Pavan, C.-T. Pham, F.~Taccone, and R.~Walther.
\newblock {GAIA} - a unified framework for sediment transport and bed evolution in rivers, coastal seas and transitional waters in the {TELEMAC-MASCARET} modelling system.
\newblock \emph{Environ. Model. Softw.}, 159:\penalty0 105544, 2023.
\newblock ISSN 1364-8152.
\newblock \doi{10.1016/j.envsoft.2022.105544}.

\bibitem[{Van der Wegen} and Jaffe(2013)]{Vanderwegen_2013}
V.~{Van der Wegen} and B.~Jaffe.
\newblock Towards a probabilistic assessment of process-based, morphodynamic models.
\newblock \emph{Coast. Eng.}, 75:\penalty0 52 -- 63, 2013.
\newblock \doi{10.1016/j.coastaleng.2013.01.009}.

\bibitem[van Rijn(1984)]{van_rijn_1984}
L.~van Rijn.
\newblock Sediment transport, part i: Bed load transport.
\newblock \emph{J. Hydraul. Eng.}, 110\penalty0 (10):\penalty0 1431--1456, 1984.
\newblock \doi{10.1061/(ASCE)0733-9429(1984)110:10(1431)}.

\bibitem[{van Rijn} et~al.(2003){van Rijn}, Walstra, Grasmeijer, Sutherland, Pan, and Sierra]{Van_Rijn_2003}
L.~{van Rijn}, D.~Walstra, B.~Grasmeijer, J.~Sutherland, S.~Pan, and J.~Sierra.
\newblock The predictability of cross-shore bed evolution of sandy beaches at the time scale of storms and seasons using process-based profile models.
\newblock \emph{Coast. Eng.}, 47\penalty0 (3):\penalty0 295--327, 2003.
\newblock ISSN 0378-3839.
\newblock \doi{10.1016/S0378-3839(02)00120-5}.

\bibitem[Villaret et~al.(2016)Villaret, Kopmann, Wyncoll, Riehme, Merkel, and Naumann]{Villaret_2016}
C.~Villaret, R.~Kopmann, D.~Wyncoll, J.~Riehme, U.~Merkel, and U.~Naumann.
\newblock First-order uncertainty analysis using {A}lgorithmic {D}ifferentiation of morphodynamic models.
\newblock \emph{Comput. Geosci.}, 90:\penalty0 144 -- 151, 2016.
\newblock \doi{10.1016/j.cageo.2015.10.012}.

\bibitem[Walstra et~al.(2004)Walstra, {Van Ormondt}, and Roelvink]{Walstra_2004}
D.~Walstra, M.~{Van Ormondt}, and J.~Roelvink.
\newblock Shoreface {N}ourishment {S}cenarios, 2004.
\newblock Technical report at \url{https://resolver.tudelft.nl/uuid:42bea38b-34f6-49fb-895c-d162c1bee193}.

\bibitem[Wilson(1966)]{Wilson_1966}
K.~Wilson.
\newblock Bed-{L}oad {T}ransport at {H}igh {S}hear {S}tress.
\newblock \emph{J. Hydraul. Div.}, 92\penalty0 (6):\penalty0 49--59, 1966.
\newblock \doi{10.1061/JYCEAJ.0001562}.

\bibitem[Wong and Parker(2006{\natexlab{a}})]{Won_2006b}
M.~Wong and G.~Parker.
\newblock Reanalysis and {C}orrection of {B}ed-{L}oad {R}elation of {M}eyer-{P}eter and {M}\"uller {U}sing {T}heir {O}wn {D}atabase.
\newblock \emph{J. Hydraul. Eng.}, 132\penalty0 (11):\penalty0 1159--1168, 2006{\natexlab{a}}.
\newblock \doi{10.1061/(ASCE)0733-9429(2006)132:11(1159)}.

\bibitem[Wong and Parker(2006{\natexlab{b}})]{Wong_2006a}
M.~Wong and G.~Parker.
\newblock Flume experiments with tracer stones under bedload transport, 2006{\natexlab{b}}.
\newblock Paper presented at the 4th IAHR Symposium on River, Coastal, and Estuarine Morphodynamics, Urbana, Illinois, USA, October, 2005.

\bibitem[Wu and Chen(2009)]{Wu_2009}
F.-C. Wu and C.~Chen.
\newblock Bayesian {U}pdating of {P}arameters for a {S}ediment {E}ntrainment {M}odel via {M}arkov {C}hain {M}onte {C}arlo.
\newblock \emph{J. Hydraul. Eng.}, 135\penalty0 (1):\penalty0 22--37, 2009.
\newblock \doi{10.1061/(ASCE)0733-9429(2009)135:1(22)}.

\bibitem[Yossef et~al.(2016)Yossef, Becker, and Deak]{Yossef_2016}
M.~Yossef, A.~Becker, and G.~Deak.
\newblock Modelling large scale and long term morphological response to engineering interventions at river bifurcation, 2016.
\newblock Paper presented at River Flow, Iowa, USA, 11--14 July 2016.

\bibitem[Zanke and Roland(2020)]{Zanke_2020}
U.~Zanke and A.~Roland.
\newblock Sediment bed-load transport: A standardized notation.
\newblock \emph{Geosciences}, 10\penalty0 (9), 2020.
\newblock ISSN 2076-3263.
\newblock \doi{10.3390/geosciences10090368}.

\bibitem[Zech et~al.(2008)Zech, {Soares-Frazão}, Spinewine, and {Le Grelle}]{Zech_2008}
Y.~Zech, S.~{Soares-Frazão}, B.~Spinewine, and N.~{Le Grelle}.
\newblock Dam-break induced sediment movement: Experimental approaches and numerical modelling.
\newblock \emph{J. Hydraul. Res.}, 46\penalty0 (2):\penalty0 176--190, 2008.
\newblock \doi{10.1080/00221686.2008.9521854}.

\bibitem[Zhu et~al.(2017)Zhu, Li, Zhang, and Duan]{Zhu_2017}
L.~Zhu, X.~Li, C.~Zhang, and Z.~Duan.
\newblock Pollutants’ {R}elease, {R}edistribution and {R}emediation of {B}lack {S}melly {R}iver {S}ediment {B}ased on {R}e-{S}uspension and {D}eep {A}eration of {S}ediment.
\newblock \emph{Int. J. Environ. Res. Public Health.}, 14\penalty0 (4), 2017.
\newblock \doi{10.3390/ijerph14040374}.

\end{thebibliography}

\end{document}